\documentclass[11pt,a4paper]{article}
\usepackage{jheppub}
\usepackage{amsmath,amssymb}
\usepackage{tcolorbox}
\usepackage{tikz}
\usepackage[compat=1.1.0]{tikz-feynman}
\usepackage{tikz-cd}
\usepackage{ctable}

\newenvironment{env}{%
\noindent\ignorespaces%
}{%
}%

\begin{document}

\begin{titlepage}

\vspace*{1.0cm}

\begin{center}
\textbf{\LARGE Kerr-Schild Double Field Theory  \\  \vskip0.3cm and Classical Double Copy}
\end{center}
\vspace{1.0cm}

\centerline{
\textsc{\large Kanghoon Lee,} $^{a}$%
\footnote{kanghoon.lee1@gmail.com} 
}

\vspace{0.6cm}

\begin{center}
${}^a${\it Fields, Gravity \& Strings @ {\rm CTPU}, Institute for Basic Science, \\
55, Expo-ro, Yuseong-gu, Daejeon 34126, \rm Korea}
\end{center}

\vspace{2cm}
\centerline{\bf Abstract}
\begin{centerline}
\noindent
The Kerr-Schild (KS) formalism is a powerful tool for constructing exact solutions in general relativity. In this paper, we present a generalization of the conventional KS formalism to double field theory (DFT) and supergravities. We introduce a generalized KS ansatz for the generalized metric in terms of a pair of null vectors. Applying this ansatz to the equations of motion of DFT, we construct the generalized KS field equation. While the generalized KS equations are quadratic in the fields, we show that it is possible to find solutions by considering linear equations only. Furthermore, we construct a Killing spinor equation under the generalized KS ansatz. Based on this formalism, we show that the classical double copy structure, which represents solutions of the Einstein equation in terms of solutions of the Maxwell equation, can be extended to the entire massless string NS-NS sector. We propose a supersymmetric classical double copy which shows that solutions of the Killing spinor equation can be realized in terms of solutions of the BPS equation of the supersymmetric Maxwell theory.

\end{centerline}

\thispagestyle{empty}

\end{titlepage}

\setcounter{footnote}{0}

\tableofcontents

\section{Introduction}
 The Kerr-Schild (KS) formalism \cite{Kerr:1963ud,KS,Debney:1969zz} has been a powerful tool for constructing exact solutions of the vacuum Einstein equations since the construction of the rotating black hole solution \cite{Kerr:1963ud}. The KS metric ansatz incorporates a geodesic null congruence $\ell$ in a background spacetime $\tilde{g}$, which plays a central role in the formalism by reducing Einstein's field equation to a set of linear differential equations. The original KS formalism describes an algebraically special vacuum spacetime in a flat background, but it is extended to arbitrary on-shell backgrounds with nontrivial matter fields, such as electromagnetism and perfect fluids (See \cite{Stephani:2003tm} for a detailed review of the KS formalism and original references).

One of the interesting applications of the KS formalism is the so called classical double copy \cite{Monteiro:2014cda}. It states that a class of solutions of the Einstein field equations can be represented by solutions of the Maxwell or linearized Yang-Mills equations, and it is examined in various examples \cite{Luna:2015paa,Ridgway:2015fdl,Luna:2016due,Goldberger:2016iau,Goldberger:2017frp,Carrillo-Gonzalez:2017iyj,Luna:2017dtq,Goldberger:2017vcg,Chester:2017vcz,Goldberger:2017ogt,Li:2018qap,Ilderton:2018lsf}. As is well known, gravity scattering amplitudes can be represented by the square of Yang-Mills scattering amplitudes in accordance with the so-called BCJ double copy relation \cite{Bern:2008qj,Bern:2010ue,Bern:2010yg}, which is inherited from the KLT relation \cite{Kawai:1985xq} in closed string theory scattering amplitudes. Since the BCJ relation holds in tree level amplitudes \cite{Stieberger:2009hq,Tye:2010dd,BjerrumBohr:2010zs,Feng:2010my,Mafra:2011kj,Monteiro:2011pc}, it is natural to deduce its extension to the level of the classical Lagrangian and the equations of motion \cite{Ananth:2007zy,Tolotti:2013caa,Bern:2010yg,Anastasiou:2014qba,Cardoso:2016ngt,Anastasiou:2016csv,Borsten:2017jpt}. From the classical double copy relation, the geodesic null vector $\ell$ in the KS ansatz is identified with the Maxwell field $A_{\mu}$. According to the BCJ relation, scattering amplitudes of the entire massless NS-NS sector can be represented by a product of two Yang-Mills scattering amplitudes. However, the classical double copy is available for the Einstein field equation only, since it is not possible to describe the Kalb-Ramond field using the single null congruence $\ell$ in the conventional KS formalism. So it is desirable to develop an alternative framework to describe the full classical double copy.

Double field theory(DFT) \cite{Siegel:1993th,Siegel:1993xq,Hull:2009mi,Hull:2009zb,Hohm:2010jy,Hohm:2010pp} is a string low energy effective field theory with manifest $\mathit{O}(d,d)$ T-duality. The manifest $\mathit{O}(d,d)$ invariance requires doubling spacetime dimensions, however, an additional constraint, the so-called section condition, should be imposed for it to be a consistent field theory. It provides a unified geometric framework for the entire closed string massless NS-NS sector encoded in DFT field variables \cite{DFTGeom1,DFTGeom2,DFTGeom3,DFTGeom4}: the generalized metric $\mathcal{H}_{MN}$ and the DFT dilaton $d$. As in general relativity, the field equations are given in terms of curvatures only: the generalized Ricci scalar and tensor. One of the crucial features of DFT is the doubled local structure group. The maximal compact subgroup of $\mathit{O}(d,d)$ that includes the Lorentz group is given by $\mathit{O}(1,d-1)_{L} \times \mathit{O}(1,d-1)_{R}$. Each local Lorentz group is originated from the left or right sector of the closed string, and this structure fits well with the double copy relation \cite{Cheung:2016say,Cheung:2017kzx}.

The aim of this paper is twofold. First, we formulate a generalization of the conventional Kerr-Schild formalism to DFT and supergravities. We introduce a generalized KS ansatz for the generalized metric in terms of a pair of null vectors satisfying the so-called generalized geodesic equation. We construct the corresponding equations of motion by substituting the generalized KS ansatz into the DFT field equations. The field equations are expressed in terms of the supergravity fields. Interestingly, unlike the conventional KS formalism in GR, the generalized KS field equations are quadratic in the fields, due to the presence of the DFT dilaton. However, we show that the generalized KS field equations can be reduced to linear equations and the field equations can be solved by considering the linear equations only. We also construct the Killing spinor equations under the generalized KS ansatz.

Second, based on the generalized KS formalism, we extend the classical double copy by including the entire massless NS-NS sector. Assuming the geometry admits a Killing vector, we derive two independent Maxwell equations by contracting a Killing vector with the generalized KS field equation. This implies that solutions of the field equation under the generalized KS ansatz can be represented by solutions of the Maxwell equation. We also identify the linearized bi-adjoint scalar field \cite{Monteiro:2011pc,Cachazo:2013iea} in a similar way and check the so-called zeroth copy relation \cite{Monteiro:2014cda}. We establish a supersymmetric generalization of the classical double copy which represent solutions of the Killing spinor equation in DFT and supergravities \cite{SUSYDFT1,SUSYDFT2,SUSYDFT3,SUSYDFT4} in terms of solutions of the BPS equations in supersymmetric Maxwell theory.

This paper is organized as follows. In section 2, we review the properties of the linearized perturbation of the generalized metric and define the generalized Kerr-Schild ansatz using the properties. We express the corresponding metric and Kalb-Ramond field ansatz as well as local frame fields in terms of the null vectors. We show that T-duality maps a generalized KS ansatz to another generalized KS ansatz. In section 3, we construct generalized KS equations in a flat background. Even though the equations are quadratic in the fields, we show that the equations can be solved by considering only linear equations. We establish the classical double copy for the entire massless NS-NS sector by extending the conventional one in GR. In section 4, we extend the results of section 3 and construct the generalized KS equations in an arbitrary on-shell background. The structure of the generalized KS equations is the same as in the flat background case, which makes it possible to solve the generalized KS equations by considering linear equations alone. In section 5, Killing spinor equations are considered under the generalized KS ansatz. We show that the Killing spinor equations lead to the BPS equation of the supersymmetric Maxwell theory. We end in section 6 by considering some examples for generalized KS formalism.

\section{Generalized Kerr-Schild ansatz in DFT} \label{sec.2}
In general relativity (GR) the Kerr-Schild ansatz is a minimal extension of linear perturbation around a background metric $\tilde{g}$. It is achieved by introducing a null vector $\ell$, and the ansatz is given by  
\begin{equation}
  g_{\mu\nu} = \tilde{g}_{\mu\nu} + \kappa \varphi \ell_{\mu} \ell_{\nu}\,,\qquad \mu,\nu,\cdots = 0,1,\cdots d-1\,,
\label{}\end{equation}
where $\kappa$ is an expansion parameter. The main advantage of the Kerr-Schild ansatz is that it preserves some features of the linearized perturbation. The form of the inverse metric and its determinant are 
\begin{equation}
  g^{\mu\nu} = \tilde{g}^{\mu\nu} - \kappa \varphi \ell^{\mu} \ell^{\nu}\,, \qquad \det(g) = \det(\tilde{g})\,.
\label{}\end{equation}
Furthermore, if we assume $\ell$ is geodesic, $\ell^{\mu}\nabla_{\mu} \ell_{\nu} =0$, the Einstein equation reduces to a linear equation\footnote{See appendix \ref{ReviewKS} for a review of the Kerr-Schild formalism in general relativity.}.

In this section we review the properties of the linearized perturbation of the generalized metric and introduce a generalized KS ansatz for the generalized metric in terms of a pair of null vectors. We introduce generalized frame fields in DFT and the corresponding generalized KS ansatz. We parametrize the generalized KS ansatz in terms of the supergravity fields. Finally, we investigate Buscher's rule and show that the form of the generalized Kerr-Schild ansatz is preserved under T-duality. 

\subsection{Linear perturbations of generalized metric}
The field content of DFT consists of the generalized metric ${\cal H}_{MN}$ and the DFT dilaton $d$. The generalized metric is a symmetric $\mathit{O}(d,d)$ tensor satisfying $\mathit{O}(d,d)$ constraint
\begin{equation}
  {\cal H}_{MN} {\cal J}^{NP} {\cal H}_{PQ} = {\cal J}_{MQ}\,,
\label{HJH}\end{equation}
where $M,N,P\cdots $ are $\mathit{O}(d,d)$ vector indices, and ${\cal J}_{MN}$ is the $\mathit{O}(d,d)$ metric with an off-diagonal form
\begin{equation}
  {\cal J}_{MN} = \begin{pmatrix} 0&\delta^{\mu}{}_{\nu} \\ \delta_{\mu}{}^{\nu} & 0\end{pmatrix}\,.
\label{}\end{equation}
It raises and lowers the $\mathit{O}(d,d)$ vector indices. Solving the $\mathit{O}(d,d)$ constraint \eqref{HJH}, one can find a parametrization of $\mathcal{H}$  in terms of the metric $g_{\mu\nu}$ and the Kalb-Ramond field $B_{\mu\nu}$
\begin{equation}
  {\cal H}_{MN} = \begin{pmatrix} g^{\mu\nu} & - g^{\mu\rho} B_{\rho\nu} \\ B_{\mu\rho} g^{\rho\nu} &~~g_{\mu\nu} - B_{\mu\rho} g^{\rho\sigma} B_{\sigma\nu} \end{pmatrix}\,.
\label{backgroundH}\end{equation}

Before discussing the generalization of the conventional KS ansatz to DFT, it is worthwhile to analyze the properties of linear perturbations of the generalized metric. We decompose the generalized metric ${\cal H}$ into the background part plus a small perturbation $\hat{\gamma}$,
\begin{equation}
  {\cal H}_{MN} = {\cal H}_{0}{}_{MN} + \hat{\gamma}_{MN}\,, \qquad |\hat{\gamma}_{MN}| \ll 1\,.
\label{Hdecomp}\end{equation}
where ${\cal H}_{0}$ is a background generalized metric satisfying the $\mathit{O}(d,d)$ constraint \eqref{HJH}. By substituting \eqref{Hdecomp} into \eqref{HJH}, we obtain a constraint on $\hat{\gamma}$. If we truncate the higher order terms and keep the linear terms, we have
\begin{equation}
  {\cal H}_{0} {\cal J}\hat{\gamma} + \hat{\gamma} {\cal J} {\cal H}_{0} = 0\,.
\label{linearHJH}\end{equation}

Since the background $\mathit{O}(d,d)$ constraint can be written as $\mathcal{H}_{0}{}^{M}{}_{N} \mathcal{H}_{0}^{N}{}_{P} = \delta^{M}{}_{P}$, it defines a chirality in the $\mathit{O}(d,d)$ vector representation and the corresponding background projection operators 
\begin{equation}
\begin{aligned}
  P_{0} &= \frac{1}{2} \big({\cal J} + {\cal H}_{0}\big)\,,& \qquad \bar{P}_{0} &= \frac{1}{2} \big({\cal J} - {\cal H}_{0}\big)	\,.
\end{aligned}\label{}
\end{equation}
These satisfy the following defining properties for projection operators:
\begin{equation}
  P^{2}_{0} = P_{0}\,, \qquad \bar{P}^{2}_{0} = \bar{P}_{0}\,, \qquad P_{0} \bar{P}_{0} = \bar{P}_{0} P_{0} = 0\,.
\label{}\end{equation}
The $P_{0}$ and $\bar{P}_{0}$ project $\mathit{O}(d,d)$ vector representation into chiral and antichiral subsectors respectively. By means of the complete relation, $\delta^{M}{}_{N} = P_{0}{}^{M}{}_{N} + \bar{P}_{0}{}^{M}{}_{N}$, $\hat{\gamma}$ is decomposed into three independent parts according to the background chiralities:
\begin{equation}
  \hat{\gamma} = \gamma + \bar{\gamma} + E + E^{t}
\label{gamma}\end{equation}
where $\gamma$ and $\bar{\gamma}$ are chiral and anti-chiral parts respectively, and $E$ is the mixed chirality part
\begin{equation}
\begin{aligned}
  \gamma = P_{0}\hat{\gamma} P_{0} \,,
  \qquad
  \bar{\gamma} = \bar{P}_{0}\hat{\gamma} \bar{P}_{0} \,,
  \qquad 
  E = P_0 \hat{\gamma} \bar{P}_0\,.
\end{aligned}\label{}
\end{equation}

The key feature of the linear perturbation of $\mathcal{H}$ is that both chiral and anti-chiral parts vanish
\begin{equation}
  \gamma = 0\,, \qquad \bar{\gamma} = 0\,.
\label{linearVanish}\end{equation}
One can easily check this by substituting \eqref{gamma} to \eqref{linearHJH}. Hence only the mixed chirality part contributes to the linear perturbation of $\mathcal{H}$, $\hat{\gamma} = E+ E^{t}$. Solving (\ref{linearVanish}), one can find the parametrization of $\hat{\gamma}$ in terms of $d$-dimensional fields 
\begin{equation}
  \hat{\gamma} = \begin{pmatrix} - \tilde{g}^{-1}h\tilde{g}^{-1} & \tilde{g}^{-1} h \tilde{g}^{-1} \tilde{B} - \tilde{g}^{-1} b \\ - \tilde{B} \tilde{g}^{-1} h \tilde{g}^{-1} + b \tilde{g}^{-1} & \quad h + \tilde{B} \tilde{g}^{-1} h \tilde{g}^{-1} \tilde{B} -b \tilde{g}^{-1} \tilde{B} - \tilde{B} \tilde{g}^{-1} b\end{pmatrix}\,,
\label{linearE}\end{equation}
where $\tilde{g}$ and $\tilde{B}$ are the background metric and Kalb-Ramond field, and $h$ and $b$ are infinitesimal symmetric and antisymmetric tensors respectively. This is consistent with the linearized perturbation of the generalized metric (\ref{backgroundH}) under the following expansion of $g$ and $B$  
\begin{equation}
  g_{\mu\nu} = \tilde{g}_{\mu\nu} + h_{\mu\nu} \,, \qquad B_{\mu\nu} = \tilde{B}_{\mu\nu} + b_{\mu\nu}\,.
\label{}\end{equation}

\subsection{Generalized Kerr-Schild ansatz}
In the above we investigated linear perturbation of the generalized metric. We showed that only the mixed chirality part $E$ contributes to the small fluctuation $\hat{\gamma}$. Following the conventional Kerr-Schild ansatz, we now assume that $\hat{\gamma}$ is a finite perturbation. It means that \eqref{Hdecomp} is no longer a linearized approximation, but an exact relation. In addition, we require \eqref{linearHJH} to hold even for finite $\hat{\gamma}$ in order to keep some properties of the linear perturbation. This implies $\hat{\gamma}$ must be a nilpotent matrix
\begin{equation}
  \hat{\gamma} {\cal J} \hat{\gamma} = 0\,.
\label{nil}\end{equation}
As we have seen in (\ref{linearVanish}) this leads to $\gamma=\bar{\gamma}=0$, and $E$ contributes to $\hat{\gamma}$ only. Thus the nilpotency condition of $\hat{\gamma}$ reads
\begin{equation}
  E_{MN} E^{t}{}^{N}{}_{P} = E^{t}{}_{MN} E^{N}{}_{P} = 0\,.
\label{nil}\end{equation}
By definition $E$ satisfies the following chirality conditions:
\begin{equation}
  E_{MN} = P_{0MP} E^{P}{}_{N} = E_{MP} \bar{P}_{0}{}^{P}{}_{N}\,.
\label{ChiralityE}\end{equation}

Now we want to represent $E$ as a product of two vectors. Any $2d\times 2d$ matrix $E$ with rank $n$ can be recast in terms of $n$-pairs of some $\mathit{O}(d,d)$ vectors $K_{M}^{a}$ and $\bar{K}_{M}^{a}$
\begin{equation}
\begin{aligned}
  E_{MN} = \sum_{a=1}^{n} \varphi^{a} K^{a}_{M} \bar{K}^{a}_{N}\,,
\end{aligned}\label{}
\end{equation}
where $\varphi^{a}$ are scalar functions. For the purpose of generalizing the conventional Kerr-Schild ansatz (\ref{KerrShild}), we set  $n=1$ throughout this paper
\begin{equation}
  E_{MN} = \varphi K_{M} \bar{K}^{t}{}_{N}\,.
\label{EMN}\end{equation}
Even in the case $n>1$, we can always describe $E$ using a single pair of vectors by absorbing the rest of them, $K^{a}$ and $\bar{K}^{a}$ for $a\geq2$, into the background generalized metric $\mathcal{H}_{0}$
\begin{equation}
\begin{aligned}
  \mathcal{H}_{MN} &= \mathcal{H}_{0MN} + \sum_{a=1}^{n} \varphi^{a} \big( K_{M}^{a} \bar{K}_{N}^{a} + K_{N}^{a} \bar{K}_{M}^{a}\big)\,,
  \\
  &= \mathcal{H}'_{0MN} + \varphi^{1} \big(K^{1}_{M} \bar{K}^{1}{}_{N}+K^{1}_{M} \bar{K}^{1}{}_{N}\big)	\,.
\end{aligned}\label{}
\end{equation}

Let us analyze the properties of $K$ and $\bar{K}$. The fact that $E$ is a nilpotent matrix implies $K$ and $\bar{K}$ are null vectors 
\begin{equation}
  K_{M} K^{M} = 0\,, \qquad \bar{K}_{M} \bar{K}^{M} = 0\,.
\label{null}\end{equation}
From the chirality condition on $E$ in (\ref{ChiralityE}), we impose background chiralities on $K^{M}$ and $\bar{K}^{M}$
\begin{equation}
  P_{0MN} K^{N} = K_{M}\,, \qquad \bar{P}_{0MN} \bar{K}^{N}= \bar{K}_{M}\,, \qquad K_{M} \bar{K}^{M} = 0\,.
\label{PKP}\end{equation}
Taking all the results together, the generalized metric can be written as
\begin{equation}
\begin{aligned}
  {\cal H}_{MN} &= {\cal H}_{0 MN} + \kappa \varphi \big(K_{M} \bar{K}_{N} + \bar{K}_{M} K_{N}\big)\,.
\end{aligned}\label{genKS}
\end{equation}
We refer this form as the \emph{generalized Kerr-Schild ansatz}. Here $\kappa$ is just a formal expansion parameter to aid in computation and does not have any physical meaning. This ansatz satisfies the $\mathit{O}(d,d)$ constraint (\ref{HJH}) automatically without any approximation or truncation. 

One can define the projection operators with respect to the total generalized metric $\mathcal{H}_{MN}$ \eqref{genKS},
\begin{equation}
  P_{MN} = \mathcal{J}_{MN} + \mathcal{H}_{MN}\,, \qquad \bar{P}_{MN} = \mathcal{J}_{MN} - \mathcal{H}_{MN}\,.
\label{}\end{equation}
The background chirality condition  \eqref{PKP} still holds even if we replace $P_{0}$ and $\bar{P}_{0}$ by $P$ and $\bar{P}$ due to  \eqref{null} and \eqref{PKP},
\begin{equation}
  P_{MN} K^{N} = K_{M}\,, \qquad \bar{P}_{MN} \bar{K}^{N}= \bar{K}_{M}\,.
\label{}\end{equation}

It is illuminating to compare with the conventional Kerr-Schild ansatz. We parametrize the $2d$-dimensional null vectors $K$ and $\bar{K}$ in terms of the $d$-dimensional vectors $l^{\mu}$ and $k_{\mu}$
\begin{equation}
  K_{M} = \frac{1}{\sqrt{2}}\begin{pmatrix} l^{\mu} \\ k_{\mu}\end{pmatrix}\,, \qquad \bar{K}_{M} = \frac{1}{\sqrt{2}}\begin{pmatrix} \bar{l}^{\mu} \\ \bar{k}_{\mu}\end{pmatrix}\,.
\label{paraK}\end{equation}
The background projection operators $P_{0}$ and $\bar{P}_{0}$ are also parametrized in terms of the background metric $\tilde{g}$ and two-form gauge field $\tilde{B}$
\begin{equation}
\begin{aligned}
  P_{0} &= \frac{1}{2}\begin{pmatrix} \tilde{g}^{-1} & \mathbf{1}_{d}  - \tilde{g}^{-1} \tilde{B} \\ \mathbf{1}_{d} + \tilde{B} \tilde{g}^{-1} & \tilde{g} - \tilde{B} \tilde{g}^{-1} \tilde{B} \end{pmatrix}\,, 
  \\
   \bar{P}_{0} &= \frac{1}{2}\begin{pmatrix} - \tilde{g}^{-1} & \mathbf{1}_{d}  + \tilde{g}^{-1} \tilde{B} \\ \mathbf{1}_{d} - \tilde{B} \tilde{g}^{-1} & - \tilde{g} + \tilde{B} \tilde{g}^{-1} \tilde{B} \end{pmatrix}\,,	
\end{aligned}\label{}
\end{equation}
where $\mathbf{1}_{d}$ is the $d$-dimensional identity matrix.
Substituting the above parametrizations into (\ref{PKP}), we get a relation between $k_{\mu}$ and $l^{\mu}$ ($\bar{k}_{\mu}$ and $\bar{l}^{\mu}$ as well)
\begin{equation}
  k_{\mu} = \big( \tilde{g}_{\mu\nu} + \tilde{B}_{\mu\nu} \big)l^{\nu} \,, \qquad \bar{k}_{\mu} = \big( -\tilde{g}_{\mu\nu} + \tilde{B}_{\mu\nu} \big)\bar{l}^{\nu} \,.
\label{relK}\end{equation}
Then $K^{M}$ and $\bar{K}^{M}$ are parametrized by $l^{\mu}$ and $\bar{l}^{\mu}$
\begin{equation}
  K_{M} = \frac{1}{\sqrt{2}}\begin{pmatrix} l^{\mu}\\ (\tilde{g} + \tilde{B})_{\mu\nu} l^{\nu} \end{pmatrix}\,, \qquad  \bar{K}_{M} = \frac{1}{\sqrt{2}}\begin{pmatrix} \bar{l}^{\mu}\\ (-\tilde{g} + \tilde{B})_{\mu\nu} \bar{l}^{\nu} \end{pmatrix}\,.
\label{paraKbK}\end{equation}
Upon substituting this result into the null condition of $K$ and $\bar{K}$ (\ref{null}), one finds that $l^{\mu}$ and $\bar{l}^{\mu}$ are also null vectors
\begin{equation}
\begin{aligned}
  l^{\mu} \tilde{g}_{\mu\nu} l^{\nu}= l^{\mu} l_{\mu} = 0\,, 
  \qquad
  \bar{l}^{\mu} \tilde{g}_{\mu\nu} \bar{l}^{\nu}= \bar{l}^{\mu} \bar{l}_{\mu}=0\,,
\end{aligned}\label{}
\end{equation}
however, $l^{\mu}$ and $\bar{l}^{\mu}$ do not have to be orthogonal to each other, $l^{\mu} \bar{l}_{\mu} \neq 0$. Here we used the background metric $\tilde{g}$ for raising and lowering the indices. Note that $l_{\mu} g^{\mu\nu} \bar{l}_{\nu} \neq l_{\mu} \tilde{g}^{\mu\nu} \bar{l}_{\nu}$ unlike the conventional Kerr-Schild ansatz. 

Consequently, $\hat{\gamma}$ is parametrized in terms of $l$ and $\bar{l}$ 
\begin{equation}
\begin{aligned}
  2K_{(M}\bar{K}_{N)} &= \begin{pmatrix} l^{(\mu} \bar{l}^{\nu)} & ~\frac{1}{2}\big(l^{\mu} \bar{k}_{\nu}+ \bar{l}^{\mu} k_{\nu}\big)\\ \frac{1}{2}\big(k_{\mu} \bar{l}^{\nu} + \bar{k}_{\mu} l^{\nu}\big)& k_{(\mu} \bar{k}_{\nu)} \end{pmatrix}	\,,
  \\
  &=\begin{pmatrix} l^{(\mu} \bar{l}^{\nu)} & ~\frac{1}{2}\big( \bar{l}^{\mu} l_{\nu}-l^{\mu}\bar{l}_{\nu} - (l^{\mu} \bar{l}^{\rho}+\bar{l}^{\mu} l^{\rho})\tilde{B}_{\rho\nu}\big)\\ \frac{1}{2}\big( l_{\mu} \bar{l}^{\nu} - \bar{l}_{\mu} l^{\nu}+ \tilde{B}_{ \mu\rho} (l^{\rho} \bar{l}^{\nu}+ \bar{l}^{\rho} l^{\nu})\big) & (l + \tilde{B} l)_{(\mu}(-\bar{l} + \tilde{B}\bar{l})_{\nu)} \end{pmatrix}	\,,
\end{aligned}\label{}
\end{equation}
where $(\pm l+\tilde{B}l)_{\mu} = \pm l_{\mu}+B_{\mu\nu}l^{\nu}$. Upon the parametrization of $\mathcal{H}$ and the generalized KS ansatz, one can read off the corresponding metric and Kalb-Ramond field 
\begin{tcolorbox}
\begin{equation}
\begin{aligned}
  (g^{-1})^{\mu\nu} &= (\tilde{g}^{-1})^{\mu\nu} + \kappa \varphi l^{(\mu} \bar{l}^{\nu)} \,,
  \\
  g_{\mu\nu} &= \tilde{g}_{\mu\nu} - \frac{\kappa \varphi }{1+\frac{1}{2}\kappa\varphi(l\cdot \bar{l})} l_{(\mu} \bar{l}_{\nu)} \,, 
  \\
  B_{\mu\nu} &= \tilde{B}_{\mu\nu} + \frac{\kappa\varphi}{1+ \frac{1}{2}\kappa \varphi (l\cdot \bar{l}) } l_{[\mu}\bar{l}_{\nu]} \,,
\end{aligned}\label{genKS_gb}
\end{equation}
where $l \cdot \bar{l} = l^{\mu} \tilde{g}_{\mu\nu} \bar{l}^{\nu}$. 
\end{tcolorbox}
If we identify $l^{\mu}$ and $\bar{l}^{\mu}$ and ignore the $\tilde{B}$, \eqref{genKS_gb} reduces to the conventional Kerr-Schild ansatz
\begin{equation}
\begin{aligned}
    g^{\mu\nu} & = \tilde{g}^{\mu\nu} + \kappa \varphi l^{\mu}l^{\nu}\,,
    \\
    g_{\mu\nu} &= \tilde{g}_{\mu\nu} - \kappa \varphi l_{\mu}l_{\nu}\,.
\end{aligned}\label{}
\end{equation}

The determinant of the metric $g$ can be computed by using Sylvester's determinant theorem. If we introduce $L$ and $\tilde{L}$,
\begin{equation}
  L_{\mu}{}^{a} =\begin{pmatrix} l_{\mu} & \bar{l}_{\mu}\end{pmatrix}\,,\qquad \tilde{L}^{a}{}_{\mu} = \begin{pmatrix} \bar{l}_{\mu} \\ l_{\mu}\end{pmatrix}\,,\qquad a =1,2\,,
\label{}\end{equation}
we can write $g_{\mu\nu}$ in terms of $L_{\mu}$ and $\tilde{L}_{\mu}$
\begin{equation}
  g_{\mu\nu} = \tilde{g}_{\mu\nu} - \frac{\kappa\varphi}{1+ \frac{1}{2}\kappa \varphi (l\cdot \bar{l}) } \sum_{a=1}^{2} L_{\mu}{}^{a}\tilde{L}^{a}{}_{\nu}\,.
\label{}\end{equation}
According to the Sylvester's identity one can compute  
\begin{equation}
  \det g = (\det \tilde{g}) \Big(1+\frac{1}{2}\kappa\varphi(l\cdot \bar{l})\Big)^{-2}\,.
\label{det}\end{equation}

It is interesting to note that the metric and the Kalb-Ramond field are not linear in $\kappa$, unlike the generalized KS ansatz \eqref{genKS}. However, as we will see later, the equations of motion are linear with respect to $\kappa$. Furthermore, one can define the $B$-field by antisymmetrizing these two null vectors, which is not available in the conventional Kerr-Schild ansatz.

\subsection{Double vielbein}
One of the most distinctive feature of DFT is the doubled local Lorentz group, $\mathit{O}(1,d-1)_{L} \times \mathit{O}(1,d-1)_{R}$ (c.f. $\mathit{O}(1,d-1) $ in Riemannain geometry). This structure is inherited from the left and right sector decomposition of the closed string. Each of the local Lorentz groups of the left and right sectors corresponds to $\mathit{O}(1,d-1)_{L}$ and $\mathit{O}(1,d-1)_{R}$ respectively. 

Now we construct the generalized frame fields, or double-vielbeins with respect to the generalized KS ansats. We introduce a background double-vielbeins $V_{0}{}_{M}{}^{m}$ and $\bar{V}_{0}{}_{M}{}^{\bar{m}}$ satisfying  \eqref{defV}, which is the defining condition of the double-vielbeins. Here $m,n,p,\cdots$ and $\bar{m},\bar{n},\bar{p},\cdots$ are generalized frame indices for $\mathit{O}(1,d-1)_{L}$ and $\mathit{O}(1,d-1)_{R}$ respectively\footnote{We use the sign convention in \cite{DFTGeom2}}. The background projection operators are denoted by 
\begin{equation}
  V_{0}{}_{M}{}^{m} V^{t}_{0}{}_{mN} = P_{0}{}_{MN}\,, \qquad \bar{V}_{0}{}_{M}{}^{\bar{m}} \bar{V}^{t}_{0}{}_{\bar{m}N} = -\bar{P}_{0}{}_{MN}\,.
\label{}\end{equation}
$V_{0}$ and $\bar{V}_{0}$ are parametrized by
\begin{equation}
  V_{0}= \frac{1}{\sqrt{2}} \begin{pmatrix} \tilde{e}^{-1} \\ \tilde{e} + \tilde{B}\tilde{e}^{-1} \end{pmatrix}\,, \qquad \bar{V}_{0} = \frac{1}{\sqrt{2}} \begin{pmatrix} \tilde{\bar{e}}^{-1} \\ -\tilde{\bar{e}} + \tilde{B}\tilde{\bar{e}}^{-1} \end{pmatrix}\,,
\label{}\end{equation}
where $\tilde{e}$ and $\tilde{\bar{e}}$ are background vielbeins corresponding to the same background metric
\begin{equation}
  \tilde{e}_{\mu}{}^{m} \tilde{e}_{m\nu} = \tilde{\bar{e}}_{\mu}{}^{\bar{m}} \tilde{\bar{e}}_{\bar{m}\nu} = \tilde{g}_{\mu\nu}\,. 
\label{}\end{equation}

Now let us consider the generalized Kerr-Schild ansatz for the double vielbeins $V_{M}{}^{m}$ and $\bar{V}_{M}{}^{\bar{m}}$. Similar to the generalized Kerr-Schild ansatz of ${\cal H}$, we divide the double-vielbein into the background part and its deviations $v_{M}{}^{m}$ and $\bar{v}_{M}{}^{\bar{m}}$,
\begin{equation}
\begin{aligned}
    V_{M}{}^{m} &= V_{0}{}_{M}{}^{m} + \kappa v_{M}{}^{m}\,,
  \\
  \bar{V}_{M}{}^{m} &= \bar{V}_{0}{}_{M}{}^{m} + \kappa\bar{v}_{M}{}^{\bar{m}}\,.
\end{aligned}\label{}
\end{equation}
Substituting the above expansion into (\ref{defV}), we find explicit expressions of $v_{M}{}^{m}$ and $\bar{v}_{M}{}^{\bar{m}}$ as follows:
\begin{equation}
  v_{M}{}^{m} = \frac{1}{2} \varphi\bar{K}_{M} K^{N} V_{0}{}_{N}{}^{m}\,, \qquad \bar{v}_{M}{}^{\bar{m}} = - \frac{1}{2}\varphi K_{M} \bar{K}^{N} \bar{V}_{0}{}_{N}{}^{\bar{m}}\,.
\label{gKS_DV}\end{equation}
Note that  the parametrization of $K$ and $\bar{K}$ in \eqref{paraKbK} can be denoted by using $V^{(0)}{}_{M}{}^{m}$ and $\bar{V}^{(0)}{}_{M}{}^{\bar{m}}$
\begin{equation}
  K_{M} = V_{0}{}_{M}{}^{m} \tilde{e}_{m\mu} l^{\mu} = V_{0}{}_{M}{}^{m} l_{m} \,, \qquad \bar{K}_{M} = \bar{V}_{0}{}_{M}{}^{\bar{m}} \tilde{\bar{e}}_{\bar{m}\mu}\bar{l}^{\mu} =\bar{V}_{0}{}_{M}{}^{\bar{m}}\bar{l}_{\bar{m}}\,.
\label{Kl}\end{equation}
Conversely, $l_{m}$ and $\bar{l}_{\bar{m}}$ are represented as
\begin{equation}
  l_{m} = V_{0}{}_{m}{}^{M}K_{M} := K_{m}\,, \qquad \bar{l}_{\bar{m}}= -\bar{V}_{0}{}_{\bar{m}}{}^{M}\bar{K}_{M} := -\bar{K}_{\bar{m}} \,.
\label{}\end{equation}
Upon this relation, we can parametrize the $v_{M}{}^{m}$ and $\bar{v}_{M}{}^{\bar{m}}$
\begin{equation}
\begin{aligned}
  v_{M}{}^{m} &= \frac{1}{2} \varphi\bar{V}_{0}{}_{M}{}^{\bar{m}} \bar{l}_{\bar{m}} l^{m} = \frac{\varphi}{2\sqrt{2}}\begin{pmatrix} \bar{l}^{\mu}l^{m} \\ (-\tilde{g}+\tilde{B})_{\mu\nu}\bar{l}^{\nu} l^{m} \end{pmatrix} \,,
  \\
  \bar{v}_{M}{}^{\bar{m}} &= \frac{1}{2}\varphi V_{0}{}_{M}{}^{m} l_{m} \bar{l}^{\bar{m}} = \frac{\varphi}{2\sqrt{2}}\begin{pmatrix} l^{\mu}\bar{l}^{\bar{m}} \\ (\tilde{g}+\tilde{B})_{\mu\nu}l^{\nu} \bar{l}^{\bar{m}} \end{pmatrix}\,.
\end{aligned}\label{}
\end{equation}
Due to the null property of $l^{\mu}$ and $\bar{l}^{\mu}$, $v$ and $\bar{v}$ satisfy
\begin{equation}
\begin{aligned}
  (v^{t})_{m}{}^{M} v_{Mn} &= 0\,, \qquad (\bar{v}^{t})_{\bar{m}}{}^{M} \bar{v}_{M\bar{n}} = 0\,, \qquad (v^{t})_{m}{}^{M} \bar{v}_{M\bar{n}} = 0\,,
  \\
  v_{M m} (v^{t}){}^{m}{}_{N} &= 0\,,\qquad \bar{v}_{M \bar{m}} (\bar{v}^{t}){}^{\bar{m}}{}_{N} = 0\,,
\end{aligned}\label{}
\end{equation}
Then the generalized Kerr-Schild ansatz for the vielbeins associated with the metric ansatz in \eqref{genKS_gb} are given by
\begin{equation}
\begin{aligned}
  e^{\mu}{}_{m} &= \tilde{e}^{\mu}{}_{m} + \frac{1}{2} \kappa\varphi \bar{l}^{\mu} l^{\nu}\tilde{e}_{\nu m}\,,\qquad e_{\mu}{}^{m} = \tilde{e}_{\mu}{}^{m} - \frac{\kappa\varphi}{2+\kappa\varphi (l\cdot \bar{l})} l_{\mu} \bar{l}^{\nu} \tilde{e}_{\nu}{}^{m}\,,
  \\
  \bar{e}^{\mu}{}_{\bar{m}} &= \tilde{\bar{e}}^{\mu}{}_{\bar{m}} + \frac{1}{2} \kappa\varphi l^{\mu} \bar{l}^{\nu}\tilde{\bar{e}}_{\nu \bar{m}}\,,\qquad \bar{e}_{\mu}{}^{\bar{m}} = \tilde{\bar{e}}_{\mu}{}^{\bar{m}} - \frac{\kappa\varphi}{2+\kappa\varphi (l\cdot \bar{l})} \bar{l}_{\mu} l^{\nu} \tilde{\bar{e}}_{\nu}{}^{\bar{m}}\,.
\end{aligned}\label{}
\end{equation}

\subsection{T-duality}
We now investigate T-duality transformation of the generalized Kerr-Schild ansatz. By definition, the generalized metric is covariant under the $\mathit{O}(d,d)$ transformation
\begin{equation}
  \mathcal{H}_{MN}' = \mathcal{O}_{M}{}^{P} \mathcal{O}_{N}{}^{Q} \mathcal{H}_{PQ} \,,
\label{T-dual}\end{equation}
where $\mathcal{O}\in \mathit{O}(d,d)$. Since it is a linear transformation, the form of the generalized Kerr-Schild ansatz should be preserved 
\begin{equation}
  \mathcal{H}'_{MN} = \mathcal{H}'_{0MN} + \kappa\varphi \big(K'_{M} \bar{K}'_{N} + K'_{N} \bar{K}'_{M} \big)\,,
\label{ODD_gKS}\end{equation}
where
\begin{equation}
  \mathcal{H}'_{0MN} = \mathcal{O}_{M}{}^{P} \mathcal{O}_{N}{}^{Q} \mathcal{H}_{0PQ}\,,
\label{OddH}\end{equation}
and
\begin{equation}
  K'_{M} = \mathcal{O}_{M}{}^{N} K_{N}\,, \qquad \bar{K}'_{M} = \mathcal{O}_{M}{}^{N} \bar{K}_{N}\,.
\label{OddK}\end{equation}

Taking the isometry direction to be $z$, we set $\mathcal{O}$ as
\begin{equation}
  \mathcal{O}_{M}{}^{N} =\left( \begin{array} { c c} { \delta^{ \mu }{}_{ \nu } - \delta^{\mu}{}_{z} \delta^{z}{}_{ \nu } } & { \delta^{\mu}{}_{z} \delta^{ \nu}{}_{z} } \\ { \delta_{\mu}{}^{z} \delta^{z}{}_{ \nu } }  & { \delta_{\mu}{}^{ \nu }  - \delta{\mu}{}^{z} \delta_{z}{}^{\nu} } \end{array} \right) = \begin{pmatrix} \delta^{i}{}_{j} & 0 & 0 & 0 \\ 0 & 0 & 0 & 1 \\ 0 & 0 & \delta_{i}{}^{j} & 0 \\ 0 & 1 & 0 & 0 \end{pmatrix}\,.
\label{}\end{equation}
From \eqref{OddH} and \eqref{OddK} one can read off Buscher's rule for the background metric and the Kalb-Ramond field 
\begin{equation}
\begin{aligned}
&\tilde{g}'_{zz} = \frac{1}{\tilde{g}_{zz}} \,,\qquad \qquad  \tilde{g}'_{iz} = \frac{\tilde{B}_{iz}}{\tilde{g}_{zz}} \,, \qquad \qquad  \tilde{B}'_{iz} = \frac{\tilde{g}_{iz}}{\tilde{g}_{zz}}\,,
\\
&\tilde{g}'_{ij} = \tilde{g}_{ij} - \frac{\tilde{g}_{iz}\tilde{g}_{jz} - \tilde{B}_{iz}\tilde{B}_{jz}}{\tilde{g}_{zz}}\,, \qquad \tilde{B}'_{ij} = \tilde{B}_{ij} - \frac{\tilde{B}_{iz}\tilde{g}_{jz} - \tilde{g}_{iz}\tilde{B}_{jz}}{\tilde{g}_{zz}} \,,
\end{aligned}\label{Buscher_gB}
\end{equation}
and for the null vectors $l$ and $\bar{l}$  
\begin{equation}
\begin{aligned}
  l'{}^{i} &= l^{i} \,, \qquad l'^{z} = \big(\tilde{g}+\tilde{B}\big){}_{z\mu} l^{\mu}\,,
  \\
  \bar{l}'{}^{i} &= \bar{l}^{i} \,, \qquad \bar{l}'^{z} = -\big(\tilde{g} -\tilde{B}\big){}_{z\mu} \bar{l}^{\mu}\,.
\end{aligned}\label{Buscher_null}
\end{equation}
Note that indices of the new null vectors $l'^{\mu}$ and $\bar{l}'^{\mu}$ are raised and lowered by new background metric, $\tilde{g}'$. Explicit calculations give components of $l'_{\mu}$ and $\bar{l}'_{\mu}$:
\begin{equation}
\begin{aligned}
  l'_{i} &= l_{i} - \frac{(\tilde{g}_{iz}-\tilde{B}_{iz})l_{z}}{\tilde{g}_{zz}}\,,\qquad l'_{z} = \frac{l_{z}}{\tilde{g}_{zz}}\,,
  \\
  \bar{l}'_{i} &= \bar{l}_{i} - \frac{(\tilde{g}_{iz}+\tilde{B}_{iz})\bar{l}_{z}}{\tilde{g}_{zz}}\,,\qquad \bar{l}'_{z} = -\frac{\bar{l}_{z}}{\tilde{g}_{zz}}\,.
\end{aligned}\label{Buscher_null}
\end{equation}
It is straightforward to show that $l'$ and $\bar{l}'$ remains as null vectors with respect to the new background metric $\tilde{g}'$,   $l'^{\mu}\tilde{g}'_{\mu\nu} l'^{\nu} = 0\,$ and $\bar{l}'^{\mu}\tilde{g}'_{\mu\nu} \bar{l}'^{\nu} = 0$. However, $l'^{\mu} \tilde{g}'_{\mu\nu} \bar{l}'^{\nu} $ is not preserved and changes to
\begin{equation}
  l'^{\mu} \tilde{g}'_{\mu\nu} \bar{l}'^{\nu} = l^{\mu}\tilde{g}_{\mu\nu} \bar{l}^{\nu} -\frac{2 l_{z} \bar{l}_{z}}{\tilde{g}_{zz}}\,.
\label{}\end{equation}
One may show that the Buscher's rule for $g_{\mu\nu}$ and $B_{\mu\nu}$ can be rewritten by the generalized KS ansatz in terms of the transformed fields
\begin{equation}
\begin{aligned}
  g'_{\mu\nu} = \tilde{g}'_{\mu\nu} - \frac{\kappa\varphi}{1+\frac{1}{2} \kappa\varphi (l'\cdot \bar{l}')} l'_{(\mu} \bar{l}'_{\nu)}\,,
  \\
  B'_{\mu\nu} = \tilde{B}'_{\mu\nu} + \frac{\kappa\varphi}{1+\frac{1}{2} \kappa\varphi (l'\cdot \bar{l}')} l'_{[\mu} \bar{l}'_{\nu]}\,.
\end{aligned}\label{}
\end{equation}

Therefrore, T-duality preserves the generalized KS ansatz form and maps a generalized KS spacetime to another one.

\section{Field equations in a flat background}
In this section we analyze DFT field equations by applying the generalized Kerr-Schild ansatz in a flat background. Similar to GR, field equations of DFT are given by geometric quantities in doubled space: the generalized Ricci tensor and scalar. Unlike the conventional KS case, DFT field equations are quadratic in $\kappa$, however, we show that the field equations can be reduced to linear equations. We also discuss how the classical double copy is realized in the generalized KS formalism.

\subsection{Constructing field equations}
In a flat background, the background generalized metric and dilaton are given by
\begin{equation}
  {\cal H}_{0}{}_{MN} = \begin{pmatrix} \eta^{\mu\nu} & 0 \\ 0 & \eta_{\mu\nu} \end{pmatrix}\,, \qquad d_{0} = \mbox{constant}\,,
\label{}\end{equation}
where $\eta_{\mu\nu}$ is a $d$-dimensional Minkowskian metric. We choose a coordinate system in which the flat metric become a constant such that $\eta=\mbox{diag}(-1,1,1,\cdots,1)$. As we have seen in the previous section,  the generalized KS ansatz is linear in $\kappa$. On the other hand, there is no restriction on DFT dilaton $d$, and it does not have to be linear in general
\begin{equation}
  d = d_{0} + \kappa f\,, \qquad f=\sum_{n=0}^{\infty} f^{(n)} \kappa^{n}\,.
\label{perturb_f}\end{equation}
The null and orthogonality conditions of $K^{M}$ and $\bar{K}^{M}$ lead to a significant simplification in the computation. For instance one can find useful identities
\begin{equation}
\begin{aligned}
    K^{M} \partial_{N} K_{M} = 0\,, \qquad \bar{K}^{M} \partial_{N} \bar{K}_{M} = 0\,, 
    \\
    K^{M} \partial_{N} \bar{K}_{M} = 0\,, \qquad \bar{K}^{M} \partial_{N} K_{M} = 0\,.
\end{aligned}\label{kdk}
\end{equation}

In DFT, local gauge symmetry is given by the generalized Lie derivative, which is a manifest $\mathit{O}(d,d)$ covariant combination of a diffeomorphism and a one-form gauge transformation of Kalb-Ramond field
\begin{equation}
\begin{aligned}
  \hat{\mathcal{L}}_{X} \mathcal{H}_{MN} &=X^{P} \partial_{P} \mathcal{H}_{MN} + \big(\partial_{M}X^{P} - \partial^{P}X_{M} \big) \mathcal{H}_{PN}+ \big(\partial_{N}X^{P} - \partial^{P}X_{N} \big) \mathcal{H}_{MP}\,,
    \\
  \hat{\mathcal{L}}_{X} \big(e^{-2d}\big) &= \partial_{M}\big(X^{M} e^{-2d}\big)\,,
\end{aligned}\label{}
\end{equation}
where $X^{M}$ is a $\mathit{O}(d,d)$ covariant gauge parameter combining the diffeomorphism parameter $\delta x^{\mu}$ and the one-form gauge parameter $\Lambda_{\mu}$, $X^{M}= \{\Lambda_{\mu}\,, \delta x^{\mu}\}$. The covariant derivative $\nabla_{M}$ and the corresponding DFT connection $\Gamma_{M}$ are defined in \cite{DFTGeom3}. Substituting the generalized KS ansatz and \eqref{perturb_f} into the DFT connection (\ref{conn}), we get \footnote{In Appendix \ref{App.1}, we briefly summarize the semi-covariant approach of DFT.}
\begin{equation}
\begin{aligned}
  \Gamma_{PMN} &= \partial_{P}\big(\varphi K_{[M}\bar{K}_{N]}\big) - \partial_{J}\big(\varphi K_{P}\bar{K}_{[M}\big) \bar{P}_{0}{}^{J}{}_{N]}  - \partial_{J}\big(\varphi \bar{K}_{P}K_{[M}\big) P_{0}{}^{J}{}_{N]} 
  \\
  &\quad -\frac{1}{2} \varphi \Big(K_{P}K_{[M}\bar{K}^{J}\partial_{|J|}\big(\varphi\bar{K}_{N]}\big)+\bar{K}_{P}\bar{K}_{[M} k^{J}\partial_{|J|}\big(\varphi K_{N]}\big)\Big)
  \\
  &\quad- \frac{4}{d-1} \big(P_{0}{}_{P[M}P_{0}{}_{N]}{}^{J}+ \bar{P}_{0}{}_{P[M}\bar{P}_{0}{}_{N]}{}^{J} +\varphi K_{P}\bar{K}_{[M}{\cal H}_{0}{}_{N]}{}^{J} +\varphi \bar{K}_{P}K_{[M}{\cal H}_{0}{}_{N]}{}^{J}\big) \partial_{J} f
  \\
  &\quad +\frac{1}{d-1} \Big(P_{0}{}_{P[M} K_{N]}  \partial_{J}\big(\varphi\bar{K}^{J})- \bar{P}_{0}{}_{P[M}\bar{K}_{N]}\partial_{J}\big(\varphi k^{J}\big)\Big)\,.
\end{aligned}\label{}
\end{equation}
It is straightforward to check that the DFT connection satisfies the following identities from (\ref{kdk})
\begin{equation}
  K^{P}\Gamma_{PMN} \bar{K}^{N} = 0\,, \qquad \bar{K}^{P}\Gamma_{PMN} K^{N} = 0\,,\qquad  \Gamma^{P}{}_{PM} K^M = \Gamma^{P}{}_{PM} \bar{K}^M = 0 \,.
\label{Gammakk}\end{equation}

Using the DFT connection, one can introduce the generalized curvature tensor $S_{MN}$ and scalar $S$, which are defined in (\ref{generalizedcurvaturescalar}). The DFT field equations are given by generalized curvatures as in GR
\begin{equation}
\begin{aligned}
  S &= 0\,,
  \qquad\qquad 
  S_{MN} &= 0\,.
\end{aligned}\label{EOM}
\end{equation}
 One can get a consistency condition by contracting $K^{M}$ and $\bar{K}^{M}$ with $\mathcal{S}_{MN}$,
\begin{equation}
\begin{aligned}
    K^{K} \bar{K}^{L}S_{KL} &= 2K^{K} \bar{K}^{L} \partial_{K}\partial_{L}{f} -\frac{1}{2}\varphi K^{K} K^{L} \partial_{K}{\bar{K}_{M}} \partial_{L}{\bar{K}^{M}}
    \\
    &\quad +\frac{1}{2}\varphi \bar{K}^{K} \bar{K}^{L}\partial_{K}{K_{M}} \partial_{L}{K^{M}} = 0\,.	
\end{aligned}\label{consistency}
\end{equation}
Analogous to the geodesic condition \eqref{geodesic} in the conventional Kerr-Schild formalism, we require 
\begin{equation}
  K^{M} \nabla_{M} \bar{K}_{N} = K^{M} \partial_{M} \bar{K}_{N}=0\,, \qquad \bar{K}^{M} \nabla_{M} K_{N} = \bar{K}^{M} \partial_{M}K_{N}=0\,, 
\label{geodesic1}\end{equation}
and 
\begin{equation}
  K^{K} \bar{K}^{L} \nabla_{K}\partial_{L}{f} = K^{K} \bar{K}^{L} \partial_{K}\partial_{L}{f} = 0\,,
\label{geodesic2}\end{equation}
where $\nabla_{M}$ is the semi-covariant derivative defined in terms of the DFT connection $\Gamma_{MNP}$ in \eqref{conn}. As the name suggests, it is not covariant under the generalized Lie derivative by itself, but it should be combined with projection operators to make it covariant
\begin{equation}
  \bar{P}_{M}{}^{P} P_{N}{}^{Q} \nabla_{P} V_{Q}\,, \qquad P_{M}{}^{P} \bar{P}_{N}{}^{Q} \nabla_{P} V_{Q}\,, 
  \qquad
  P^{PQ} \nabla_{P} V_{Q}\,, \qquad \bar{P}^{PQ} \nabla_{P} V_{Q}\,.
\label{}\end{equation}
Thus, one can check that \eqref{geodesic1} and \eqref{geodesic2} are covariant under the generalized Lie derivative as well as the $\mathit{O}(d,d)$ transformation. Substituting the parametrization of $K^{M}$ and $\bar{K}^{M}$ into \eqref{paraKbK}, they yield the following $d$-dimensional expressions:
\begin{equation}
\begin{aligned}
  l^{\mu} \partial_{\mu} \bar{l}_{\nu} &= 0\,,&\qquad \bar{l}^{\mu} \partial_{\mu} l_{\nu} &= 0\,,&\qquad  l^{\mu}\bar{l}^{\nu} \partial_{\mu} \partial_{\nu}f &= 0\,, 
\end{aligned}\label{gengeo}
\end{equation}
where we have used the section condition, namely
\begin{equation}
  \partial_{M} = \begin{pmatrix} \tilde{\partial}^{\mu} \\ \partial_{\mu} \end{pmatrix} = \begin{pmatrix} 0 \\ \partial_{\mu} \end{pmatrix}\,.
\label{}\end{equation}
If we identify $l^{\mu}$ and $\bar{l}^{\mu}$ and ignore $f$, (\ref{gengeo}) reduces to the conventional geodesic equation (\ref{geodesic}). Hence we will denote \eqref{geodesic1} and \eqref{geodesic2} as {\it generalized geodesic equations}, though the notion of geodesic is not obvious in the doubled spacetime point of view.

By applying the generalized Kerr-Schild ansatz to \eqref{EOM}, we can represent the DFT field equations in terms of the null vectors $K^{M}$ and $\bar{K}^{M}$
\begin{equation}
\begin{aligned}
  S_{KL} &= \kappa\Big[-\frac{1}{2}{\cal H}_{0}^{M N} \partial_{M}\partial_{N}\big(\varphi K_{(K} \bar{K}_{L)}\big)+\partial_{M} \partial_{N}\big(\varphi K^{N} \bar{K}_{(K} P_0{}_{L)}{}^{M} - \varphi K_{(K} \bar{K}^{N} \bar{P}_0{}_{L)}{}^{M}\big)
  \\
  &\quad +4 P_{0}{}_{(K}{}^{M} \bar{P}_{0}{}_{L)}{}^{N}\partial_{M}\partial_{N}{f} -\frac{\varphi^{2}}{4}  {\cal H}_{0}^{M N} \big( K_{K} K_{L}\partial_{M}{\bar{K}_{P}} \partial_{N}{\bar{K}^{P}} -\bar{K}_{K} \bar{K}_{L} \partial_{M}{K_{P}} \partial_{N}{K^{P}} \big) \ \Big] 
  \\
  &\quad + \kappa^{2}\Big[\ {\cal H}_{0}^{MN}\partial_{M}f \partial_{N}\big(\varphi K_{(K} \bar{K}_{L)}\big) -2 P_{0(K}{}^{M} \partial_{|M|}\big( \varphi K^{N}\bar{K}_{L)}\partial_{N}f\big)
  \\
  &\quad  + 2\bar{P}_{0(K}{}^{M}\partial_{|M|}\big( \varphi K_{L)}\bar{K}^{N}\partial_{N}f\big)-2\varphi \big(P_{0}{}_{(K}{}^{M} K_{L)}\bar{K}^{N} - \bar{P}_{0}{}_{(K}{}^{M} K^{N}\bar{K}_{L)}\big)\partial_{M}\partial_{N}f
  \\
  &\quad +\frac{\varphi}{2} \bar{K}_{K} \bar{K}_{L} K^{M} \partial_{M}\partial_{N}(K^{N}\varphi) +\frac{\varphi}{2} K_{K} K_{L} \bar{K}^{M} \partial_{M}\partial_{N}(\bar{K}^{N} {\varphi})\ \Big] = 0 \,,
\end{aligned}\label{KSgenRicci}
\end{equation}
and
\begin{equation}
\begin{aligned}
  S &= - 2\kappa \partial_{K} \partial_{L} ( \varphi K^{K} \bar{K}^{L}) + 4\kappa{\cal H}_{0}^{K L} \partial_{K}\partial_{L}{f} -4\kappa^{2} {\cal H}_{0}^{K L} \partial_{K}{f} \partial_{L}{f} = 0\,.	
\end{aligned}\label{}
\end{equation}
Using the background projection operators, one can decompose $S_{MN}$ into three independent parts according to the background chiralities:
\begin{align}
  P_{0}{}_{K}{}^{M} P_{0}{}_{L}{}^{N}S_{MN} &= -2\kappa^{2}\varphi K_{(K} \bar{K}^{M} P_{0}{}_{L)}{}^{N}\partial_{N}\partial_{M}{f} -\frac{\kappa^{2}}{4} \varphi^{2} {\cal H}_{0}^{M N} K_{K} K_{L}\partial_{M}{\bar{K}_{P}} \partial_{N}{\bar{K}^{P}} \nonumber
  \\
  & \quad + \frac{\kappa^{2}}{2}\varphi K_{K} K_{L}\bar{K}^{M} \partial_{M}\partial_{N}(\bar{K}^{N} {\varphi}) =0\,, \label{PP}
  \\
  \bar{P}_{0}{}_{K}{}^{M} \bar{P}_{0}{}_{L}{}^{N}S_{MN} &= 2 \kappa^{2}\varphi^{2} K^{M} \bar{K}_{(K} \bar{P}_{0}{}_{L)}{}^{N}\partial_{N}\partial_{M}{f} -\frac{\kappa^{2}}{4} \varphi^{2}{\cal H}_{0}^{M N}  \bar{K}_{K} \bar{K}_{L} \partial_{M} K_{P}  \partial_{N} K^{P} \nonumber
  \\
  &\quad +\frac{\kappa^{2}}{2}\varphi \bar{K}_{K} \bar{K}_{L} K^{M} \partial_{M}\partial_{N}(K^{N}\varphi)=0\,,\label{bPbP}
  \\
  P_{0}{}_{K}{}^{M} \bar{P}_{0}{}_{L}{}^{N}S_{MN} &= -\frac{\kappa}{2}{\cal H}_{0}^{M N} \partial_{M}\partial_{N}\big(\varphi K_{K} \bar{K}_{L}\big)+\kappa\partial_{M} \partial_{N}\big(\varphi K^{N} \bar{K}_{K}\big) P_0{}_{L}{}^{M}\nonumber
  \\
  &\quad  - \kappa\partial_{M}\partial_{N}\big(\varphi K_{K} \bar{K}^{N}\big) \bar{P}_0{}_{L}{}^{M} +4\kappa P_{0}{}_{K}{}^{M} \bar{P}_{0}{}_{L}{}^{N}\partial_{M}\partial_{N}{f} \nonumber
  \\
  &\quad +\kappa^{2} {\cal H}_{0}^{MN}\partial_{M}f \partial_{N}\big(\varphi K_{K} \bar{K}_{L}\big) -2\kappa^{2} P_{0K}{}^{M} \partial_{M}\big( \varphi K^{N}\bar{K}_{L}\partial_{N}f\big)\nonumber
  \\
  &\quad  + 2\kappa^{2}\bar{P}_{0K}{}^{M}\partial_{M}\big( \varphi K_{L}\bar{K}^{N}\partial_{N}f\big)=0\,. \label{PbP}
\end{align}
Interestingly, (\ref{PP}) and (\ref{bPbP}) are redundant equations, and these can be derived from the mixed chiral part (\ref{PbP})
\begin{equation}
\begin{aligned}
  P_{0}{}_{K}{}^{M}P_{0}{}_{L}{}^{N}S_{MN} &= -\frac{1}{2}K_{K} \bar{K}^{M} P_{0}{}_{L}{}^{P} \bar{P}_{0}{}_{M}{}^{Q}S_{PQ}\,,
  \\
  \bar{P}_{0}{}_{K}{}^{M}\bar{P}_{0}{}_{L}{}^{N}S_{MN} &= -\frac{1}{2}\bar{K}_{K} K^{M} P_{0}{}_{L}{}^{P} \bar{P}_{0}{}_{M}{}^{Q}S_{PQ}\,.
\end{aligned}\label{}
\end{equation}
Therefore the independent field equations are as follows:
\begin{equation}
\begin{aligned}
  {\cal R} = - 2\kappa \partial_{K} \partial_{L} ( \varphi K^{K} \bar{K}^{L}) + 4\kappa{\cal H}_{0}^{K L} \partial_{K}\partial_{L}{f} -4\kappa^{2} {\cal H}_{0}^{K L} \partial_{K}{f} \partial_{L}{f} = 0\,,
\end{aligned}\label{EoM1}
\end{equation}
and
\begin{equation}
\begin{aligned}
  {\cal R}_{KL} &= \kappa\Big[-\frac{1}{2}{\cal H}_{0}^{M N} \partial_{M}\partial_{N}\big(\varphi K_{K} \bar{K}_{L}\big)+\partial_{M} \partial_{N}\big(\varphi K^{N} \bar{K}_{L} P_0{}_{K}{}^{M} - \varphi K_{K} \bar{K}^{N} \bar{P}_0{}_{L}{}^{M}\big)
  \\
  &\qquad +4 P_{0}{}_{K}{}^{M} \bar{P}_{0}{}_{L}{}^{N}\partial_{M}\partial_{N}{f}\ \Big] +\kappa^{2} \Big[\ {\cal H}_{0}^{MN}\partial_{M}f \partial_{N}\big(\varphi K_{K} \bar{K}_{L}\big)   
  \\
  &\qquad-2 P_{0K}{}^{M} \partial_{M}\big( \varphi K^{N}\bar{K}_{L}\partial_{N}f\big) + 2\bar{P}_{0L}{}^{M}\partial_{M}\big( \varphi K_{K}\bar{K}^{N}\partial_{N}f\big) \ \Big]=0\,.
\end{aligned}\label{EoM2}
\end{equation}
where ${\cal R}_{KL} = P_{0}{}_{K}{}^{M} \bar{P}_{0}{}_{L}{}^{N}S_{MN}$ and we replaced the notation $S$ into ${\cal R}$.  

Equations \eqref{EoM1} and \eqref{EoM2} are written in terms of $\mathit{O}(d,d)$ vector indices for manifest $\mathit{O}(d,d)$ covariance. It is useful to formulate the theory, however, the $\mathit{O}(d,d)$ indices are not convenient when we solve the equations of motion due to the redundancies in the components of $\mathcal{R}_{MN}$. Among the $d(2d+1)$ components, only  $d^{2}$ components are independent. Furthermore, the components of $K_M$ and $\bar{K}_M$ are not independent as well, and they are written in terms of $l^{m}$ and $\bar{l}^{\bar{m}}$ through \eqref{Kl}. Thus we recast the field equations in terms of the generalized frame indices by using the background double vielbeins, $V_{0M}{}^{m}$ and $\bar{V}_{0M}{}^{\bar{m}}$
\begin{equation}
\begin{aligned}
  {\cal R} &= \kappa\Big[ - \partial_{m}\partial_{\bar{n}} \big(\varphi l^{m} \bar{l}^{\bar{n}}\big) + 4 \Box f\ \Big] -4 \kappa^{2} \partial_{m}f \partial^{m}f = 0\,,
  \\
  {\cal R}_{m\bar{n}} &= \kappa\Big[\ \frac{1}{4}\Box \big(\varphi l_{m} \bar{l}_{\bar{n}}\big) -\frac{1}{4} \partial_{m} \partial_{p} \big(\varphi l^{p} \bar{l}_{\bar{n}}\big) - \frac{1}{4} \partial_{\bar{n}} \partial_{\bar{p}} \big(\varphi l_{m} \bar{l}^{\bar{p}}\big) + \partial_{m} \partial_{\bar{n}} f \ \Big] 
  \\
  &\quad - \frac{\kappa^{2}}{2} \Big[\ \partial^{p}f \partial_{p}\big(\varphi l_{m} \bar{l}_{\bar{n}}\big) -\partial_{m}\big(\varphi l^{p}\bar{l}_{\bar{n}} \partial_{p}f\big) -\partial_{\bar{n}}\big(\varphi l_{m}\bar{l}^{\bar{p}} \partial_{\bar{p}}f\big)  \Big]= 0\,,
\end{aligned}\label{Rmbn}
\end{equation}
where $\Box = \partial_{m} \partial^{m} = \partial_{\bar{m}} \partial^{\bar{m}} = \partial_{\mu}\partial^{\mu}$. Here we have used
\begin{equation}
\begin{aligned}
    P_{MN} &= V_{Mm} V_{N}{}^{m}\,, &  \bar{P}_{MN} &= -\bar{V}_{M\bar{m}} \bar{V}_{N}{}^{\bar{m}}\,,
    \\
    V_{0}{}^{M}{}_{m} \partial_{M} &= \frac{1}{\sqrt{2}} \partial_{m} = \frac{1}{\sqrt{2}} e_{m}{}^{\mu} \partial_{\mu}\,,&\qquad\bar{V}_{0}{}^{M}{}_{\bar{m}} \partial_{M} &= \frac{1}{\sqrt{2}} \partial_{\bar{m}}= \frac{1}{\sqrt{2}} \bar{e}_{\bar{m}}{}^{\mu} \partial_{\mu}\,. 
\end{aligned}\label{}
\end{equation}
Note that ${\cal R}_{m\bar{n}}$ is not a symmetric tensor. The symmetric and antisymmetric parts correspond to the field equations of metric and Kalb-Ramond field respectively.

There are still redundancies even in the generalized frame representation, since we can take a further gauge fixing on the doubled local Lorentz group, $\mathit{O}(1,d-1)_{L} \times \mathit{O}(1,d-1)_{R} \to \mathit{O}(1,d-1)_{D}$, by identifying the two background vielbeins, $\tilde{e}_{\mu}{}^{m}$ and $\tilde{\bar{e}}_{\mu}{}^{\bar{m}}$. Thus it is convenient to rewrite the field equations in terms of $d$-dimensional vector indices $\mu,\nu,\rho \cdots$. Thus the DFT and supergravity equations of motion under the generalized Kerr-Schild ansatz read as follows:
\begin{align}
  \mathcal{R}:  &~\kappa\Big[\ \partial_{\mu}\partial_{\nu} \big(\varphi l^{\mu} \bar{l}^{\nu}\big) - 4 \Box f\ \Big] + 4 \kappa^{2} \partial_{\mu}{f}\partial^{\mu}{f}=0\,, \label{eom_dil}
\\
  \mathcal{R}_{\mu\nu}:	 &~\kappa \Big[\ \Box\big(\varphi l_{\mu}\bar{l}_{\nu}\big) - \partial^{\rho}\partial_{\mu}\big(\varphi l_{\rho}\bar{l}_{\nu}\big) - \partial^{\rho}\partial_{\nu}\big(\varphi l_{\mu}\bar{l}_{\rho}\big) + 4 \partial_{\mu} \partial_{\nu} f \ \Big] \nonumber
  	 \\
  	 &-2\kappa^{2}\Big[  \partial_{\rho}f \partial^{\rho}\big(\varphi l_{\mu} \bar{l}_{\nu}\big) - \partial_{\mu}\big(\varphi l^{\rho}\bar{l}_{\nu} \partial_{\rho} f\big) - \partial_{\nu}\big(\varphi l_{\mu}\bar{l}^{\rho} \partial_{\rho} f\big)\ \Big]=0 \,.	\label{eom_gen}
\end{align}

It is interesting that the generalized KS ansatz for $g_{\mu\nu}$ and $B_{\mu\nu}$ is not linear in $\varphi$, $l^{\mu}$ and $\bar{l}^{\mu}$, but the field equations are linear in these fields. However, unlike the conventional KS formalism in GR, the above equations are quadratic in $\kappa$ due to the presence of $f$. We will show in the following subsection that the field equations can be solved by considering only linear equations.

\subsection{Solving the field equations} \label{sec:3.2}
One of the peculiar properties of \eqref{eom_dil} and \eqref{eom_gen} is that all the $\mathcal{O}(\kappa^{2})$ terms contain $f$. In other words, we get purely linear equations in $\kappa$ when $f=0$. Using this fact, we propose two strategies depending on which field is solved first.

Let us consider an approach solving for $\kappa f$ first.  Notice that if we fix $\kappa f$ to be a specific function, the equations of motion reduce to linear equations in $\kappa$. Then we can solve the corresponding $\varphi$, $l$ and $\bar{l}$ for a given $\kappa f$ by solving the linear equations. As a preliminary step, we recast \eqref{eom_dil} as
\begin{equation}
\begin{aligned}
  \partial_{\mu}\partial_{\nu} \big(\varphi l^{\mu} \bar{l}^{\nu}\big) + 4 {\cal F}^{-1} \Box {\cal F}=0\,,
\end{aligned}\label{Neom_dil}
\end{equation}
where ${\cal F}(x) = e^{-\kappa f(x)}$.
Without loss of generality, one can divide this equation into two pieces by introducing a function $\lambda(x)$ that we should fix by hand
\begin{equation}
\begin{aligned}
    \Box {\cal F}_{\lambda} + \frac{1}{4}\lambda(x) {\cal F}_{\lambda}=0\,,
    \\
    \lambda(x) = \partial_{\mu}\partial_{\nu} \big(\varphi l^{\mu} \bar{l}^{\nu}\big)\,.
\end{aligned}\label{}
\end{equation}
Here $\mathcal{F}_{\lambda}$ denotes the solution for a given $\lambda(x)$. Choosing a suitable $\lambda(x)$, we can solve for $\mathcal{F}$ exactly, since the first equation can be thought to be an inhomogeneous massless Klein-Gordon equation. Once we solve for ${\cal F}_{\lambda}$, we may substitute the corresponding $\kappa f_{\lambda}$ into \eqref{eom_gen}, and we obtain a linear equation
\begin{equation}
\begin{aligned}
  &\Box\big(\varphi l_{\mu}\bar{l}_{\nu}\big) - \partial^{\rho}\partial_{\mu}\big(\varphi l_{\rho}\bar{l}_{\nu}\big) - \partial^{\rho}\partial_{\nu}\big(\varphi l_{\mu}\bar{l}_{\rho}\big) + 4 \partial_{\mu} \partial_{\nu} f_{\lambda} 
  \\
  	 &-2 \partial_{\rho}f_{\lambda} \partial^{\rho}\big(\varphi l_{\mu} \bar{l}_{\nu}\big) +2\partial_{\mu}\big(\varphi l^{\rho}\bar{l}_{\nu} \partial_{\rho} f_{\lambda}\big) +2\partial_{\nu}\big(\varphi l_{\mu}\bar{l}^{\rho} \partial_{\rho} f_{\lambda}\big)=0 \,.
\end{aligned}\label{eom_gen_lambda}
\end{equation}

Let's consider two simple cases.
\begin{itemize}
  \item ${\cal F}=1$ or $f=0$ 
  \\
  The simplest case is just to set $f = 0$. In terms of the supergravity dilaton, this corresponds to $e^{-2\phi} = \frac{1}{\sqrt{-g}}$.
Then the field equations reduce to
\begin{equation}
\begin{aligned}
  &\partial_{\mu}\partial_{\nu} \big(\varphi l^{\mu} \bar{l}^{\nu}\big)=0\,,
  \\
  &\Box\big(\varphi l_{\mu}\bar{l}_{\nu}\big) - \partial^{\rho}\partial_{\mu}\big(\varphi l_{\rho}\bar{l}_{\nu}\big) - \partial^{\rho}\partial_{\nu}\big(\varphi l_{\mu}\bar{l}_{\rho}\big) =0\,.
\end{aligned}\label{}
\end{equation}

\item $\lambda = \text{constant}$\\
In this case, ${\cal F}_{\lambda}$ satisfies the Klein-Gordon equation for a real massive scalar field. For example, ${\cal F}$ is represented by the Fourier integral in a plane wave basis
\begin{equation}
  {\cal F}_{\lambda}(x)= \int {\rm d}\omega \ {\rm d}^{d-1}k\ \tilde{ {\cal F}}_{\lambda}(\omega,k_{i}) \delta(\omega^{2}-|\vec{k}|^{2}-\frac{\kappa}{4}\lambda)e^{-i(\omega t + k_{i}x^{i})}\,.
\label{}\end{equation}
In particular, if $\lambda = 0$ and ${\cal F}= {\cal F}(t,x)$, there are infinitely many solutions, ${\cal F}= {\cal F}^{+}(u) + {\cal F}^{-}(v)$, where $u=x+t$ and $v=x-t$. 
\end{itemize}
This method is somewhat ad hoc, but it may capture the nonlinear effects of \eqref{eom_dil} and \eqref{eom_gen}.

Now let us consider the second strategy solving $\varphi$, $l^{\mu}$ and $\bar{l}^{\mu}$ first. Recall that the generalized KS ansatz \eqref{genKS} is linear in $\kappa$, but there is no a priori restriction on $f$ and it is expanded in $\kappa$ to arbitrary order. Inserting the expansion of $f$ \eqref{perturb_f} into \eqref{eom_dil} and \eqref{eom_gen}, we get an expansion of the DFT field equations in $\kappa$
\begin{equation}
\begin{aligned}
  {\cal R} = \sum_{n=1}^{\infty} {\cal R}^{(n)} \kappa^{n}\,,\qquad   {\cal R}_{\mu\nu} = \sum_{n=1}^{\infty} {\cal R}^{(n)}{}_{\mu\nu}\, \kappa^{n}\,.
\end{aligned}\label{}
\end{equation}
The linear order equations are 
\begin{equation}
\begin{aligned}
  &\partial_{\mu}\partial_{\nu} \big(\varphi l^{\mu} \bar{l}^{\nu}\big) - 4 \Box f^{(0)}=0\,,
\\
  & \Box\big(\varphi l_{\mu}\bar{l}_{\nu}\big) - \partial^{\rho}\partial_{\mu}\big(\varphi l_{\rho}\bar{l}_{\nu}\big) - \partial^{\rho}\partial_{\nu}\big(\varphi l_{\mu}\bar{l}_{\rho}\big) + 4 \partial_{\mu} \partial_{\nu} f^{(0)} =0	\,,
\end{aligned}\label{linearEoM_flat}
\end{equation}
and the higher order equations, $ n\geq 0$, are 
\begin{equation}
\begin{aligned}
  \Box f^{(n+1)} &= \sum_{p+q=n} \partial_{\mu}{f^{(p)}}\partial^{\mu}{f^{(q)}}\,, 
  \\
  \partial_{\mu} \partial_{\nu} f^{(n+1)} &= \frac{1}{2} \Big[\partial_{\rho}f^{(n)} \partial^{\rho}\big(\varphi l_{\mu} \bar{l}_{\nu}\big)- \partial_{\mu}\big(\varphi l^{\rho}\bar{l}_{\nu} \partial_{\rho} f^{(n)}\big) - \partial_{\nu}\big(\varphi l_{\mu}\bar{l}^{\rho} \partial_{\rho} f^{(n)}\big) \Big]\,.
\end{aligned}\label{higherEoM_flat}
\end{equation}

Note that $\varphi$, $l^{\mu}$, $\bar{l}^{\mu}$ and $f^{(0)}$ can be completely determined from the linear equations only \eqref{linearEoM_flat}, and the higher order equations define recursion relations with respect to $f^{(n)}$, for $n>0$. This means that the metric and the Kalb-Ramond field are determined from the linear equations only. In principle we can employ \eqref{higherEoM_flat} to determine $f^{(n)}$ recursively, however, it is very difficult to determine $f$ exactly by using the recursion relation. 

Let us assume that $\varphi$, $l^{\mu}$ and $\bar{l}^{\mu}$ are given by solving the linear equations. Instead of taking the approach using the recursion relations, we substitute $\varphi$, $l^{\mu}$ and $\bar{l}^{\mu}$ into \eqref{eom_gen} and \eqref{Neom_dil}. Since these equations are linear with respect to $\mathcal{F}$ and $f$, one can determine $f$ completely.

Take a look at the set of linear equations \eqref{linearEoM_flat}. If we combine the first equation and the trace of the second equation, we obtain a simple relation
\begin{equation}
\begin{aligned}
  \Box \big(\varphi l \cdot \bar{l}-4 f^{(0)}\big) =0\,.
\end{aligned}\label{}
\end{equation}
This implies that $ f^{(0)}$ is the same as $\frac{1}{4}\varphi l\cdot \bar{l} $ up to the harmonic function $H(x)$, which satisfies $\Box H = 0$,
\begin{equation}
  f^{(0)} = \frac{1}{4}\big(\varphi l\cdot \bar{l} + H(x)\big)\,.
\label{}\end{equation}
To determine the harmonic function $H(x)$, we need to impose an appropriate boundary condition on $f^{(0)}$. 
\begin{tcolorbox}
Consequently, we can find exact solutions for DFT and supergravities just by solving the following linear differential equation,
\begin{equation}
\begin{aligned}
  & \Box\big(\varphi l_{\mu}\bar{l}_{\nu}\big) - \partial^{\rho}\partial_{\mu}\big(\varphi l_{\rho}\bar{l}_{\nu}\big) - \partial^{\rho}\partial_{\nu}\big(\varphi l_{\mu}\bar{l}_{\rho}\big) + \partial_{\mu}\partial_{\nu} \big(\varphi l \cdot \bar{l}\big)+ \partial_{\mu} \partial_{\nu} H =0\,.
\end{aligned}\label{SolvEOM_flat}
\end{equation}
\end{tcolorbox}
\begin{env}
As a consistency check, one can show that if we identify the null vectors $l$ and $\bar{l}$, \eqref{SolvEOM_flat} reduces to the conventional KS equation \eqref{Ricci1} with the $\partial_{\mu}\partial_{\nu}H(x)$ term, which can be interpreted as a matter.   
\end{env}

Note that the solutions of \eqref{SolvEOM_flat} are the solutions of \eqref{eom_dil} and \eqref{eom_gen} automatically, but the converse is not true. This is because the linear equations does not capture the full nonlinear effects. However, \eqref{SolvEOM_flat} is remarkably simple compare to the supergravity equations of motion, and it provides a useful tool for finding exact solutions of supergravities.  

\subsection{Classical double copy} \label{sec:3.3}
One of the remarkable properties of gravity scattering amplitude is the double copy structure. It is a field theory generalization of the KLT relation \cite{Kawai:1985xq} which comes from the left-right sector decomposition of closed string theory. The KLT relation states that a closed string scattering amplitude can be represented by two open string scattering amplitudes.  In the field theoretic viewpoint, it is known as the BCJ double copy relation \cite{Bern:2008qj,Bern:2010ue,Bern:2010yg}, which expresses duality between the color and kinematic factor in the Yang-Mills amplitude. By interchanging the color factor and kinematic factor, the gravity amplitude can be obtained from the Yang-Mills scattering amplitude.

Recently, the double copy structure in scattering amplitude has been extended to the level of the classical equations of motion  by using the conventional KS formalism \cite{Monteiro:2014cda}. It is the so called classical double copy, and the Maxwell equation can be derived from the Einstein field equation in the framework of the conventional KS formalism (See appendix \ref{ReviewKS}). The null geodesic vector $\ell^{\mu}$ is identified as a Maxwell field, and the scalar field $\phi$ in \eqref{KerrShild} is identified as the linearized bi-adjoint scalar field. 

The KLT and BCJ relations indicate that not only the Einstein field equation, but also the field equations of entire massless NS-NS sector should be related to the Maxwell equation. However, it is not possible to describe the Kalb-Ramond field using the single null vector $\ell^{\mu}$. 

We now extend the classical double copy to include the entire massless NS-NS sector, $g_{\mu\nu}$, $B_{\mu\nu}$ and $\phi$, through the generalized Kerr-Schild formalism. First, we need to assume that the full geometry admits at least one Killing vector $\xi^{\mu}$ which satisfies
\begin{equation}
  {\cal L}_{\xi} g = 0 \,, \qquad {\cal L}_{\xi} B = 0 \,, \qquad {\cal L}_{\xi} \phi = 0 \,.
\label{isometry_flat}\end{equation}
We can locally choose a coordinate system $x^{\mu} = \{x^{i},y\}$ such that the Killing vector is a constant, $\xi^{\mu} = \partial x^{\mu}/\partial y = \delta^{\mu}_{y}$. In this coordinate system, the Lie derivative of the full metric $g$ is given by
\begin{equation}
\begin{aligned}
  {\cal L}_{\xi} g_{\mu\nu} &=\xi^{\rho} \partial_{\rho} g_{\mu\nu} =-\xi^{\rho}\partial_{\rho}\Big(\frac{\kappa \varphi }{1+\frac{1}{2}\kappa\varphi(l\cdot \bar{l})} l_{(\mu} \bar{l}_{\nu)}\Big)=0\,.
\end{aligned}\label{Lie_g}
\end{equation}
Note that the trace part of the above equation vanishes, $\xi^{\mu}\partial_{\mu} (\eta^{\mu\nu} g_{\mu\nu})=0$, and we have
\begin{equation}
  \xi^{\mu}\partial_{\mu} (\varphi l\cdot \bar{l})=0\,.
\label{Lie_trace}\end{equation}
Next, the Kalb-Ramond field satisfies
\begin{equation}
\begin{aligned}
    {\cal L}_{\xi} B_{\mu\nu} &=\xi^{\rho} \partial_{\rho} B_{\mu\nu} = -\xi^{\rho}\partial_{\rho}\Big(\frac{\kappa\varphi}{1+ \frac{1}{2}\kappa \varphi (l\cdot \bar{l}) } l_{[\mu}\bar{l}_{\nu]}\Big)=0\,.	
\end{aligned}\label{Lie_B}
\end{equation}
By combining \eqref{Lie_g}, \eqref{Lie_trace} and \eqref{Lie_B} we get
\begin{equation}
  \xi^{\rho}\partial_{\rho} \Big(\varphi l_{\mu} \bar{l}_{\nu} \Big) = 0\,.
\label{}\end{equation}
Finally, $f$ should also satisfy
\begin{equation}
  {\cal L}_{\xi} f = \xi^{\mu}\partial_{\mu} f =0\,.
\label{}\end{equation}
We also normalize $l_{\mu}$ and $\bar{l}_{\mu}$ as follows:
\begin{equation}
  \xi\cdot l = \xi\cdot \bar{l} =1\,.
\label{x}\end{equation}

By contracting the Killing vector with the generalized KS field equation in \eqref{SolvEOM_flat}, we get the so called zeroth and single copy.
\begin{itemize}
\item Single Copy

Similar to GR, one may expect that the Maxwell equation can be obtained by contracting a Killing vector with one of the indices of the generalized KS equation \eqref{SolvEOM_flat}. Since \eqref{SolvEOM_flat} is neither symmetric nor antisymmetric, two different Maxwell theories arise depending on which index is contracted with the Killing vector. 
\begin{equation}
\begin{aligned}
   & \Box (\varphi l_{\mu}) - \partial^{\rho}\partial_{\mu}(\varphi l_{\rho}) =0\,,
  \\
  & \Box (\varphi \bar{l}_{\nu}) - \partial^{\rho}\partial_{\nu}(\varphi \bar{l}_{\rho})  =0 \,.
\end{aligned}\label{cont_R}
\end{equation}
Following \cite{Monteiro:2014cda}, we define a pair of abelian gauge fields  
\begin{equation}
  A_{\mu} = \varphi l_{\mu}\,, \qquad \bar{A}_{\mu} = \varphi \bar{l}_{\mu} \,.
\label{}\end{equation}
Then \eqref{cont_R} reduces to a pair of Maxwell equations
\begin{equation}
\begin{aligned}
 &\partial^{\mu} F_{\mu\nu} = 0 \,,
 \qquad
  &\partial^{\mu} \bar{F}_{\mu\nu} = 0\,,
\end{aligned}\label{}
\end{equation}
where $F_{\mu\nu}$ and $\bar{F}_{\mu\nu}$ are the field strengths of $A_{\mu}$ and $\bar{A}_{\mu}$ respectively,
\begin{equation}
  F_{\mu\nu} = \partial_{\mu}A_{\nu} - \partial_{\nu}A_{\mu}\,, \qquad \bar{F}_{\mu\nu} = \partial_{\mu}\bar{A}_{\nu} - \partial_{\nu}\bar{A}_{\mu}\,.
\label{Field_Strength}\end{equation}

\item Zeroth Copy

Now let us contract all the free indices of \eqref{SolvEOM_flat} with $\xi^{\mu}$. We get a scalar equation
\begin{equation}
  \Box \varphi = 0 \,.
\label{zeroth}\end{equation}
Again, following the identification in \cite{Monteiro:2014cda}, the scalar field $\varphi$ can be interpreted as an abelianized version of the biadjoint scalar field $\Phi^{aa'}$ \cite{Monteiro:2011pc,Cachazo:2013iea}. 
\end{itemize}

This ensures that solutions of the generalized Kerr-Schild equation \eqref{SolvEOM_flat} can be represented by Maxwell gauge fields and a harmonic scalar function when there is an isometry. However, the converse is not true in general.

\section{Generalization to curved Backgrounds}\label{sec 4}
In the previous section, we presented DFT field equations using the generalized KS ansatz in a flat background. In this section we extend the previous flat background results to arbitrary curved backgrounds. To this end we construct doubled spin connections and generalized curvatures by using the curved background generalized Kerr-Schild ansatz. We discuss how to solve the field equations by reducing to linear equations. 

\subsection{Uplifting to general backgrounds}
Let us return to the generalized Kerr-Schild ansatz around a general background ${\cal H}_{0}$ and $d_{0}$ in \eqref{backgroundH},
\begin{equation}
    {\cal H}_{MN} = {\cal H}_{0 MN} + \kappa \varphi \big(K_{M} \bar{K}_{N} + \bar{K}_{M} K_{N}\big)\,,
\qquad d = d_{0} + \kappa f\,.
\label{CurvedGenKS}\end{equation}
The null and orthogonality conditions of $K_{M}$ and $\bar{K}_{M}$ lead the following identities which are similar to \eqref{kdk}
\begin{equation}
\begin{aligned}
    &K^{M} \nabla_{N} K_{M} = K^{M} \partial_{N} K_{M} = 0\,,& \qquad &\bar{K}^{M} \nabla_{N} \bar{K}_{M} =\bar{K}^{M} \partial_{N} \bar{K}_{M} = 0\,, 
    \\
    &K^{M} \nabla_{N} \bar{K}_{M} =K^{M} \nabla_{0N} \bar{K}_{M} = 0\,,& \qquad &\bar{K}^{M} \nabla_{N} K_{M} =\bar{K}^{M} \nabla_{0N} K_{M} = 0\,,
\end{aligned}\label{}
\end{equation}
where $\nabla_{0M}$ is the background covariant derivative 
\begin{equation}
  \nabla_{0M} = \partial_{M} + \Gamma_{0M}\,,
\label{}\end{equation}
where $\Gamma_{0MNP}$ is the background DFT connection.
One can show that the DFT connection satisfies the following identities (\textit{c.f.} \eqref{Gammakk})
\begin{equation}
\begin{aligned}
 \bar{K}^{M} \Gamma_{MNP} K^{P} &= -\bar{P}_{MN} \bar{K}^{P}\partial_{P}K^{M} + P_{N}{}^{P} K_{M} \partial_{P}\bar{K}^{M} - P_{MN} K^{P}\partial_{P}\bar{K}^{M}\,,
 \\
 K^{M} \Gamma_{MNP} \bar{K}^{P} &= - P_{NP} K^{M} \partial_{M}\bar{K}^{P} + \bar{P}^{M}{}_{N} \bar{K}_{P} \partial_{M}K^{P} - \bar{P}_{MN} \bar{K}^{P} \partial_{P} K^{M}\,,
 \\
 K^{N} \Gamma_{MNP} \bar{K}^{P} &= - K_{P} \partial_{M}\bar{K}^{P} \,,
\end{aligned}\label{identityGamma}
\end{equation}
where $P$ and $\bar{P}$ are projection operators for total $\mathcal{H}$. Using these identities, one can show that
\begin{equation}
\begin{aligned}
  \bar{K}^{M} \Gamma_{MNP} K^{P} &= \bar{K}^{M} \Gamma_{0MNP} K^{P} \,,
  \\
  K^{M} \Gamma_{MNP} \bar{K}^{P} &= K^{M} \Gamma_{0MNP} \bar{K}^{P} \,.
\end{aligned}\label{connectionID}
\end{equation}

Just as for the flat background, we obtain consistency conditions on $K^{M}$ and $\bar{K}^{M}$ by contracting them with generalized Ricci tensor, 
\begin{equation}
\begin{aligned}
  K^{K} \bar{K}^{L}S_{KL} &=  2 K^{K} \bar{K}^{L} \nabla_{K}\partial_{L}{f} -\frac{1}{2}\varphi K^{K} K^{L} \nabla_{K}{\bar{K}_{M}} \nabla_{L}{\bar{K}^{M}}
  \\
  &\quad+\frac{1}{2}\varphi \bar{K}^{K} \bar{K}^{L}\nabla_{K}{K_{M}} \nabla_{L}{K^{M}} = 0\,.
\end{aligned}\label{consistency_curved}
\end{equation}
We impose the generalized geodesic equations 
\begin{equation}
\begin{aligned}
  K^{K} \nabla_{K} \bar{K}_{L} = K^{K} \nabla_{0K} \bar{K}_{L} = 0\,, \qquad \bar{K}^{K} \nabla_{K} K_{L} = \bar{K}^{K} \nabla_{0K} K_{L} =0\,, 
\end{aligned}\label{geodesic_curved}
\end{equation}
and 
\begin{equation}
    K^{K} \bar{K}^{L} \nabla_{L}\partial_{K}{f} = K^{K} \bar{K}^{L} \nabla_{0L}\partial_{K}{f} = 0\,,
\label{geodesic_curved2}\end{equation}
where we have used \eqref{connectionID}. We may also write the generalized geodesic equation using \eqref{identityGamma}
\begin{equation}
\begin{aligned}
  	\bar{K}^{M} \nabla_{M} K_{N} &= P_{MN} \big(\bar{K}^{P}\partial_{P}K^{M} + K_{P}\partial^{M} \bar{K}^{P} - K^{P}\partial_{P}\bar{K}^{M} \big) = 0\,,
  	\\
  	K^{M} \nabla_{M} \bar{K}_{N} &= \bar{P}_{MN} \big(K^{P} \partial_{P}\bar{K}^{M} -K_{P}\partial^{M} \bar{K}^{P} - \bar{K}^{P} \partial_{P} K^{M}\big) = 0\,,
\end{aligned}\label{}
\end{equation}
and these imply a new relation
\begin{equation}
  \bar{K}^{P}\partial_{P}K^{M} + K_{P}\partial^{M} \bar{K}^{P} - K^{P}\partial_{P}\bar{K}^{M} = 0\,.
\label{}\end{equation}

Upon the parametrization of $K_M$ and $\bar{K}_M$, \eqref{geodesic_curved} and \eqref{geodesic_curved2} reduce to
\begin{equation}
\begin{aligned}
    \bar{l}^{\mu} \tilde{\triangledown}^{+}_{\mu} l^{\nu} &=0\,,&\qquad  l^{\mu} \tilde{\triangledown}^{-}_{\mu} \bar{l}^{\nu}&=0\,, \qquad   l^{\mu} \bar{l}^{\nu} \tilde{\triangledown}^{-}_{\mu}\partial_{\nu} f &= 0\,, 
\end{aligned}\label{}
\end{equation}
where $\tilde{\triangledown}_{\mu}$ is the usual covariant derivative in Riemannian geometry with respect to the background metric $\tilde{g}$, and $\tilde{\triangledown}^{\pm}_{\mu}$ is a torsionful covariant derivative which acts on an arbitrary vector field $v_{\mu}$
\begin{equation}
\begin{aligned}
  \tilde{\triangledown}^{+}_{\mu} l_{\nu} &= \tilde{\triangledown}_{\mu} l_{\nu} + \frac{1}{2} \tilde{H}_{\mu\nu\rho} l^{\rho}\,,
  \\
  \tilde{\triangledown}^{-}_{\mu} \bar{l}_{\nu} &=\tilde{\triangledown}_{\mu} l_{\nu} - \frac{1}{2} \tilde{H}_{\mu\nu\rho} l^{\rho}\,.
\end{aligned}\label{background_connection_torsion}
\end{equation}

We now introduce the double-vielbein and the associated double spin connection, $\Phi_{Mmn}$ and $\bar{\Phi}_{M\bar{m}\bar{n}}$ (See Appendix \ref{App.1}). By assuming compatibility of the double-vielbeins, $\mathcal{D}_{M} V_{N}{}^{m} = 0$ and $\mathcal{D}_{M} \bar{V}_{N}{}^{\bar{m}} = 0$, the double spin connection is represented as 
\begin{equation}
\begin{aligned}
  \Phi_{Mmn} &= V^{N}{}_{m}\nabla_{M}V_{Nn} =V^{N}{}_{m}\partial_{M}V_{Nn} + \Gamma_{MNP} V^{N}{}_{m} V^{P}{}_{n}\,,
  \\
  \bar{\Phi}_{M\bar{m}\bar{n}} &= \bar{V}^{N}{}_{\bar{m}}\nabla_{M}\bar{V}_{N\bar{n}} =\bar{V}^{N}{}_{\bar{m}}\partial_{M}\bar{V}_{N\bar{n}} + \Gamma_{MNP} \bar{V}^{N}{}_{\bar{m}} \bar{V}^{P}{}_{\bar{n}}\,.
\end{aligned}\label{}
\end{equation}
Note that the DFT spin connection is not covariant by itself under the generalized diffeomorphism. However, one can extract the following covariant objects, and these are building blocks of the theory
\begin{equation}
\begin{aligned}
  	\bar{V}^{M}{}_{\bar{p}}\Phi_{Mmn}\,, \qquad V^{M}{}_{[p}\Phi_{M mn]}\,, \qquad V^{Mm}\Phi_{Mmn}\,,
  	\\
  	V^{M}{}_{p}\bar{\Phi}_{M\bar{m}\bar{n}} \,, \qquad \bar{V}^{M}{}_{[\bar{p}}\bar{\Phi}_{M\bar{m}\bar{n}]}\,, \qquad \bar{V}^{M\bar{m}}\bar{\Phi}_{M\bar{m}\bar{n}}\,.
\end{aligned}\label{}
\end{equation}
Applying the generalized Kerr-Schild ansatz for the double-vielbeins \eqref{gKS_DV}, we find the covariant pieces of DFT spin connections : 
\begin{equation}
\begin{aligned}
  \bar{V}^{M}{}_{\bar{p}}\Phi_{Mmn} &= \bar{V}_{0}^{M}{}_{\bar{p}} \Phi_{0}{}_{Mmn} + \kappa \Big[ {\cal D}_{0}{}_{M}\big(\varphi\bar{K}_{\bar{p}} K_{[m}\big)V_{0}^{M}{}_{n]}-\frac{1}{2}\varphi \bar{K}_{\bar{p}}K^{q} V^{M}_{0}{}_{q}\Phi_{0Mmn}\Big]\,,
  \\
  V^{M}{}_{p}\bar{\Phi}_{M\bar{m}\bar{n}} &= V_{0}^{M}{}_{p}\bar{\Phi}_{0M\bar{m}\bar{n}} -\kappa \Big[{\cal D}_{0M}\big(\varphi K_{p}\bar{K}_{[\bar{m}}\big)\bar{V}_{0}^{M}{}_{\bar{n}]} + \frac{1}{2}\varphi K_{p}\bar{K}^{\bar{q}} \bar{V}_{0}^{M}{}_{\bar{q}}\bar{\Phi}_{0M\bar{m}\bar{n}}\Big]\,,
  \\
  V^{M}{}_{[p}\Phi_{M mn]} &= V_{0}^{M}{}_{[p}\Phi_{0M mn]} + \frac{1}{2} \kappa\varphi \Phi_{0M[np}K_{m]}\bar{K}^{M}\,,
  \\
  \bar{V}^{M}{}_{[\bar{p}}\bar{\Phi}_{M \bar{m}\bar{n}]} &= \bar{V}_{0}^{M}{}_{[\bar{p}}\bar{\Phi}_{0M \bar{m}\bar{n}]} - \frac{1}{2} \kappa\varphi \bar{\Phi}_{0M[\bar{n}\bar{p}}\bar{K}_{\bar{m}]}K^{M}\,,
  \\
  V^{Mm}\Phi_{Mmn} &= V_{0}^{Mm}\Phi_{0Mmn} +\frac{\kappa}{2} \Big[{\cal D}_{0M}\big(\varphi K_{n} \bar{K}^{M}\big) -\Phi_{0Mn}{}^{q}\varphi K_{q}\bar{K}^{M} -4 V_{0}^{M}{}_{n}\partial_{M}f\Big]
  \\
  & \quad - \kappa^{2} \varphi K_{n} \bar{K}^{M} \partial_{M}f
  \,,
  \\
  \bar{V}^{M\bar{m}}\bar{\Phi}_{M\bar{m}\bar{n}} &= \bar{V}_{0}^{M\bar{m}}\bar{\Phi}_{0M\bar{m}\bar{n}} + \frac{\kappa}{2} \Big[{\cal D}_{0M}\big(\varphi K^{M} \bar{K}_{\bar{n}}\big)+\bar{\Phi}_{0M\bar{n}}{}^{\bar{q}}\varphi K^{M}\bar{K}_{\bar{q}} +4 \bar{V}_{0}^{M}{}_{\bar{n}}\partial_{M}f\Big]
   \\
  & \quad - \kappa^{2} \varphi K^{M}\bar{K}_{\bar{n}}\partial_{M}f\,.
\end{aligned}\label{DFT_Connection_GKS}
\end{equation}

For later use, we represent these quantities in terms of $d$-dimensional fields. Here, all the background fields are denoted by a tilde. First, let us introduce $\tilde{D}_{\mu}$ and $\tilde{D}^{\pm}_{\mu}$ which are background covariant derivative with torsion, which acts as
\begin{equation}
\begin{aligned}
  \hat{\tilde{D}}_{\mu} T_{m\bar{n}} &= \partial_{\mu} T_{m\bar{n}} +\tilde{\omega}^{+}_{\mu m}{}^{p} T_{p\bar{n}} - \bar{\tilde{\omega}}^{-}_{\mu\bar{n}}{}^{\bar{p}} T_{m\bar{p}}\,,
  \\
  \tilde{D}_{\mu}^{\pm} T_{m\bar{n}} &= \partial_{\mu} T_{m\bar{n}} +\tilde{\omega}^{\pm}_{\mu m}{}^{p} T_{p\bar{n}} - \bar{\tilde{\omega}}^{\pm}_{\mu\bar{n}}{}^{\bar{p}} T_{m\bar{p}}\,,
\end{aligned}\label{}
\end{equation}
where the background spin connections are defined by using \eqref{background_connection_torsion}
\begin{equation}
\begin{aligned}
  \tilde{\omega}_{\mu mn} &= e^{\nu}{}_{m} \tilde{\triangledown}_{\mu}e_{\nu n}\,,& \qquad \bar{\tilde{\omega}}_{\mu \bar{m}\bar{n}} &= \bar{e}^{\nu}{}_{\bar{m}} \tilde{\triangledown}_{\mu}\bar{e}_{\nu \bar{n}}\,,
  \\
  \tilde{\omega}^{\pm}_{\mu mn} &= e^{\nu}{}_{m} \tilde{\triangledown}^{\pm}_{\mu}e_{\nu n}\,,& \qquad \bar{\tilde{\omega}}^{\pm}_{\mu \bar{m}\bar{n}} &= \bar{e}^{\nu}{}_{\bar{m}} \tilde{\triangledown}^{\pm}_{\mu}\bar{e}_{\nu \bar{n}}\,.
\end{aligned}\label{}
\end{equation}
Then, generalizing \eqref{MasterDerivative} and \eqref{rep_DFTconnection} to the background quantities, we can recast \eqref{DFT_Connection_GKS} in terms of the background $d$-dimensional supergravity fields and the null vectors $l$ and $\bar{l}$
\begin{equation}
\begin{aligned}
  \sqrt{2} \bar{V}^{M}{}_{\bar{p}}\Phi_{Mmn} &= \tilde{\omega}^{+}_{\bar{p} mn} +\kappa \Big[ \hat{\tilde{D}}_{[m} \big(\varphi l_{n]} \bar{l}_{\bar{p}} \big) + \frac{1}{2} \varphi \bar{l}_{\bar{p}} l^{q} \tilde{\omega}^{-}_{qmn} \Big] \,,
  \\
  \sqrt{2} V^{M}{}_{p}\bar{\Phi}_{M\bar{m}\bar{n}} &= - \bar{\tilde{\omega}}^{-}_{p\bar{m}\bar{n}} -\kappa \Big[\hat{\tilde{D}}_{[\bar{m}}\big(\varphi l_{p}\bar{l}_{\bar{n}]}\big) + \frac{1}{2}\varphi l_{p}\bar{l}^{\bar{q}}\tilde{\bar{\omega}}^{+}_{\bar{q}\bar{m}\bar{n}}\Big]\,,
  \\
  \sqrt{2} V^{M}{}_{[p}\Phi_{M mn]} &= \tilde{\omega}^{+}_{[pmn]} -\frac{1}{3}\tilde{H}_{mnp}+ \frac{1}{2} \kappa\varphi l_{[m}\bar{l}^{\bar{q}} \tilde{\omega}^{+}_{\bar{q}np]}\,,
  \\
  \sqrt{2}\bar{V}^{M}{}_{[\bar{p}}\bar{\Phi}_{M \bar{m}\bar{n}]} &= -\tilde{\bar{\omega}}^{-}_{[\bar{p}\bar{m}\bar{n}]} -\frac{1}{3}\tilde{H}_{\bar{m}\bar{n}\bar{p}} - \frac{1}{2} \kappa\varphi l^{q}\bar{l}_{[\bar{m}}\bar{\tilde{\omega}}^{-}_{q\bar{n}\bar{p}]}\,,
  \\
  \sqrt{2}V^{Mm}\Phi_{Mmn} &= \tilde{\omega}^{m}{}_{mn} -2 \partial_{n}\tilde{\phi} +\frac{\kappa}{2} \Big[\hat{\tilde{D}}_{\bar{p}}\big(\varphi l_{n} \bar{l}^{\bar{p}}\big) -\varphi l_{q}\bar{l}^{\bar{p}}\tilde{\omega}^{+}_{\bar{p}n}{}^{q} -4 \partial_{n}f\Big] 
  \\
  &\quad -\kappa^{2} \varphi l_{n}\bar{l}^{\bar{p}} \partial_{\bar{p}} f\,,
  \\
  \sqrt{2}\bar{V}^{M\bar{m}}\bar{\Phi}_{M\bar{m}\bar{n}} &= -\tilde{\bar{\omega}}^{\bar{m}}{}_{\bar{m}\bar{n}} +2\partial_{\bar{n}}\tilde{\phi} -\frac{\kappa}{2} \Big[\hat{\tilde{D}}_{m}\big(\varphi l^{m} \bar{l}_{\bar{n}}\big) -\varphi l^{m}\bar{l}_{\bar{q}} \tilde{\bar{\omega}}^{-}_{m\bar{n}}{}^{\bar{q}}-4 \partial_{\bar{n}}f\Big] 
  \\
  &\quad  + \kappa^{2} \varphi l^{p} \bar{l}_{\bar{n}} \partial_{p} f\,.
\end{aligned}\label{}
\end{equation}

\subsection{Equations of motion}
We now consider the DFT equations of motion with the generalized Kerr-Schild ansatz around curved backgrounds. The linear perturbation of DFT is studied by Ko et al. \cite{Ko:2015rha}. However, as we have seen in the previous section, the field equations include terms quadratic in $\kappa$, thus we need to adopt an alternative approach. Here we employ a method using the double spin connection \eqref{DFT_Connection_GKS}. Substituting \eqref{DFT_Connection_GKS} into the DFT field equations in terms of the double spin connections \eqref{generalizedcurvaturescalar}, we get
\begin{equation}
\begin{aligned}
 {\cal R} =  \kappa & \Big[\ \big({\tilde{D}}_{m}-2\partial_{m} \tilde{\phi}\big){\tilde{D}}_{\bar{n}}\big(\varphi l^{m}\bar{l}^{\bar{n}}\big) +\frac{1}{2} \tilde{H}^{pm\bar{n}} \tilde{D}_{p}\big(\varphi l_{m}\bar{l}_{\bar{n}}\big) - \frac{1}{2} \tilde{H}_{pqm} \tilde{H}^{pq}{}_{\bar{n}}\big(\varphi l^{m}\bar{l}^{\bar{n}}\big)
  \\
  &\  - 4\big( {\tilde{D}}^{m}- 2\partial^{m}\tilde{\phi}\big)\partial_{m}f  + \big(\mathcal{G}_{m \bar{n}} + 2\tilde{\cal B}_{m \bar{n}}\big) \big(\varphi l^{m}\bar{l}^{\bar{n}}\big)\Big] +4 \kappa^{2} \partial^{m}f \partial_{m}f = 0\,,	
\end{aligned}\label{}
\end{equation}
and
\begin{equation}
\begin{aligned}
 {\cal R}_{m\bar{n}}&= \frac{\kappa}{4} \Big[\ \big(\hat{\tilde{D}}_{p}-2 \partial_{p}\tilde{\phi}\big) \hat{\tilde{D}}^{p}\big(\varphi l_{m} \bar{l}_{\bar{n}}\big) -\big(\hat{\tilde{D}}^{p}-2\partial^{p}\tilde{\phi}\big) \hat{\tilde{D}}_{m}\big(\varphi l_{p}\bar{l}_{\bar{n}}\big)
   \\
  &\qquad\ - \big(\hat{\tilde{D}}^{\bar{p}}-2\partial^{\bar{p}}\tilde{\phi}\big) \hat{\tilde{D}}_{\bar{n}}\big(\varphi l_{m}\bar{l}_{\bar{p}}\big) +\tilde{H}_{p\bar{n}}{}^{\bar{q}} \hat{\tilde{D}}^{p}\big(\varphi l_{m}\bar{l}_{\bar{q}}\big) -\tilde{H}_{pm}{}^{q} \hat{\tilde{D}}^{p}\big(\varphi l_{q}\bar{l}_{\bar{n}} \big) 
  \\
  &\qquad\ +4 \hat{\tilde{D}}_{m} \partial_{\bar{n}} f  + \big(\tilde{{\cal G}}_{q m}- \tilde{{\cal B}}_{qm} \big) \big(\varphi l^{q}\bar{l}_{\bar{n}}\big) + \big(\tilde{{\cal G}}_{\bar{n}\bar{q}} - \tilde{{\cal B}}_{\bar{n}\bar{q}}\big)\big(\varphi l_{m}\bar{l}^{\bar{q}}	\big) \ \Big]
  \\
  &\quad - \frac{\kappa^{2}}{2} \Big[\ \partial^{p}f \hat{\tilde{D}}_{p}\big(\varphi l_{m} \bar{l}_{\bar{n}}\big) - \hat{\tilde{D}}_{\bar{n}}\big(\varphi l_{m} \bar{l}^{\bar{p}}\partial_{\bar{p}} f\big) -\hat{\tilde{D}}_{m}\big(\varphi l^{p} \bar{l}_{\bar{n}}\partial_{p} f\big)
  \\
  &\qquad\qquad  -\frac{1}{4} \varphi^{2} l_{m}\bar{l}_{\bar{n}} l^{p}\bar{l}^{q} \big(\tilde{{\cal G}}_{p\bar{q}}-\tilde{{\cal B}}_{p\bar{q}}\big)\Big]
   = 0\,,
\end{aligned}\label{}
\end{equation}
where $\tilde{\cal G}$ and $\tilde{\cal B}$ are the field equations of supergravity for the background metric and Kalb-Ramond field 
\begin{equation}
\begin{aligned}
  \tilde{\cal G}_{\mu\nu} &= \tilde{R}_{\mu\nu} + 2 \tilde{\triangledown}_{(\mu}\partial_{\nu)}\tilde{\phi} - \frac{1}{4} \tilde{H}_{\mu}{}^{\rho\sigma}\tilde{H}_{\nu \rho\sigma}\,,
  \\
  \tilde{\cal B}_{\mu\nu} &= -\frac{1}{2} \tilde{\triangledown}^{\rho} \tilde{H}_{\rho\mu\nu} + \partial^{\rho}\tilde{\phi} \tilde{H}_{\rho\mu\nu}\,.
\end{aligned}\label{}
\end{equation}
Since we are assuming on-shell backgrounds, the background field equations vanish, $\tilde{\mathcal{G}}=0$ and $\tilde{\mathcal{B}}=0$. 

To make these look more familiar let us rewrite the field equations in terms of $d$-dimensional vector indices, $\mu, \nu, \rho, \cdots$, by using a pair of vielbeins $\tilde{e}_{\mu}{}^{m}$ and $\bar{\tilde{e}}_{\mu}{}^{\bar{m}}$
\begin{equation}
\begin{aligned}
  {\cal R}&=\kappa\Big[\ \big(\tilde{\triangledown}_{\mu} -2\partial_{\mu}\tilde{\phi}\big)\tilde{\triangledown}_{\nu} \big(\varphi l^{\mu}\bar{l}^{\nu}\big) +\frac{1}{2} \tilde{H}^{\mu\nu\rho} \partial_{\mu}\big(\varphi l_{\nu}\bar{l}_{\rho}\big) -\frac{1}{2} \tilde{H}_{\rho\sigma\mu} \tilde{H}^{\rho\sigma}{}_{\nu} \big(\varphi l^{\mu}\bar{l}^{\nu}\big)
  \\
  & \qquad -4 \big({\tilde{\triangledown}}^{\mu}- 2\partial^{\mu}\tilde{\phi} \big)\partial_{\mu}f \ \Big] +4 \kappa^{2} \partial^{\mu}f \partial_{\mu}f = 0\,,
\end{aligned}\label{eom_curved_dil}
\end{equation}
and
\begin{equation}
\begin{aligned}
  {\cal R}_{\mu\nu}&=\frac{\kappa}{4} \Big[\ \big({\tilde{\triangledown}}_{\rho}-2 \partial_{\rho}\tilde{\phi}\big){\tilde{\triangledown}}^{\rho}\big(\varphi l_{\mu} \bar{l}_{\nu}\big) -\big({\tilde{\triangledown}}^{\rho}-2\partial^{\rho}\tilde{\phi}\big) {\tilde{\triangledown}}_{\mu}\big(\varphi l_{\rho}\bar{l}_{\nu}\big)- \big({\tilde{\triangledown}}^{\rho}-2\partial^{\rho}\tilde{\phi}\big) {\tilde{\triangledown}}_{\nu}\big(\varphi l_{\mu}\bar{l}_{\rho}\big) 
  \\
  &\qquad\ +\tilde{\triangledown}^{\rho} \tilde{H}_{\mu\nu}{}^{\sigma} \big(\varphi l_{(\rho}\bar{l}_{\sigma)}\big) -\tilde{H}_{\mu\rho}{}^{\sigma}{\tilde{\triangledown}}^{\rho} \big(\varphi l_{\sigma}\bar{l}_{\nu}\big) +\tilde{H}_{\nu\rho}{}^{\sigma} {\tilde{\triangledown}}^{\rho} \big(\varphi l_{\mu}\bar{l}_{\sigma}\big) -\tilde{H}_{ (\mu}{}^{\rho\sigma} {\tilde{\triangledown}}_{\nu)}\big(\varphi l_{\rho}\bar{l}_{\sigma}\big)
 \\
  &\qquad\ + \tilde{H}_{\mu\nu}{}^{\rho} \big(\tilde{\triangledown}^{\sigma} -2\partial^{\sigma}\tilde{\phi}\big)\big(\varphi l_{(\rho} \bar{l}_{\sigma)}\big) -\tilde{H}_{\tau( \mu}{}^{\rho} \tilde{H}^{\tau}{}_{\nu)}{}^{\sigma} \big(\varphi l_{\rho} \bar{l}_{\sigma}\big) +4 \tilde{\triangledown}^{-}_{\mu} \partial_{\nu} f \ \Big] 
  \\
  &\quad -\frac{\kappa^{2}}{2} \Big[\ \partial^{\rho}f \tilde{\triangledown}_{\rho}\big(\varphi l_{\mu} \bar{l}_{\nu}\big) -\tilde{\triangledown}_{\mu}\big(\varphi l^{\rho} \bar{l}_{\nu}\partial_{\rho}f\big) - \tilde{\triangledown}_{\nu}\big(\varphi l_{\mu} \bar{l}^{\rho}\partial_{\rho}f\big)
  \\
  &\qquad \qquad + \partial^{\rho}f \tilde{H}_{\rho\mu}{}^{\sigma} \varphi l_{\sigma} \bar{l}_{\nu} - \partial^{\rho}f \tilde{H}_{\rho\nu}{}^{\sigma} \varphi l_{\mu} \bar{l}_{\sigma} + \partial^{\rho}f \tilde{H}_{\mu\nu}{}^{\sigma} \varphi l_{(\rho}\bar{l}_{\sigma)}\ \Big]= 0\,.
\end{aligned}\label{eom_curved_gen}
\end{equation}

The structure of the field equations are the same as in the flat background case. The field equations are quadratic in $\kappa$, however, if we take $f = 0$, we get linear equations in $\kappa$. Hence we can still apply the strategies for solving equations of motion discussed in section \ref{sec:3.2}. Here we describe the second strategy only. 

Let us insert the expansion of $f$ \eqref{perturb_f} into \eqref{eom_curved_dil} and \eqref{eom_curved_gen} to expand order by order in $\kappa$
\begin{itemize}
\item linear equations
\begin{equation}
\begin{aligned}
  &\big(\tilde{\triangledown}_{\mu} -2\partial_{\mu}\tilde{\phi}\big) \tilde{\triangledown}_{\nu} \big(\varphi l^{\mu}\bar{l}^{\nu}\big) +\frac{1}{2} \tilde{H}^{\mu\nu\rho} \partial_{\mu}\big(\varphi l_{\nu}\bar{l}_{\rho}\big) -\frac{1}{2} \tilde{H}_{\rho\sigma\mu} \tilde{H}^{\rho\sigma}{}_{\nu} \big(\varphi l^{\mu}\bar{l}^{\nu}\big)
  \\
  &-4 \big({\tilde{\triangledown}}^{\mu}- 2\partial^{\mu}\tilde{\phi} \big)\partial_{\mu}f^{(0)}= 0\,,
\end{aligned}\label{linear_dil_curved}
\end{equation}
and
\begin{equation}
\begin{aligned}
& \big({\tilde{\triangledown}}_{\rho}-2 \partial_{\rho}\tilde{\phi}\big){\tilde{\triangledown}}^{\rho}\big(\varphi l_{\mu} \bar{l}_{\nu}\big) -\big({\tilde{\triangledown}}^{\rho}-2\partial^{\rho}\tilde{\phi}\big) {\tilde{\triangledown}}_{\mu}\big(\varphi l_{\rho}\bar{l}_{\nu}\big)- \big({\tilde{\triangledown}}^{\rho}-2\partial^{\rho}\tilde{\phi}\big) {\tilde{\triangledown}}_{\nu}\big(\varphi l_{\mu}\bar{l}_{\rho}\big) 
  \\
  &+\tilde{\triangledown}^{\rho} \tilde{H}_{\mu\nu}{}^{\sigma} \big(\varphi l_{(\rho}\bar{l}_{\sigma)}\big) -\tilde{H}_{\mu\rho}{}^{\sigma}{\tilde{\triangledown}}^{\rho} \big(\varphi l_{\sigma}\bar{l}_{\nu}\big) +\tilde{H}_{\nu\rho}{}^{\sigma} {\tilde{\triangledown}}^{\rho} \big(\varphi l_{\mu}\bar{l}_{\sigma}\big) -\tilde{H}_{ (\mu}{}^{\rho\sigma} {\tilde{\triangledown}}_{\nu)}\big(\varphi l_{\rho}\bar{l}_{\sigma}\big)
 \\
  &+ \tilde{H}_{\mu\nu}{}^{\rho} \big(\tilde{\triangledown}^{\sigma} -2\partial^{\sigma}\tilde{\phi}\big)\big(\varphi l_{(\rho} \bar{l}_{\sigma)}\big) -\tilde{H}_{\tau( \mu}{}^{\rho} \tilde{H}^{\tau}{}_{\nu)}{}^{\sigma} \big(\varphi l_{\rho} \bar{l}_{\sigma}\big) +4 \tilde{\triangledown}^{-}_{\mu} \partial_{\nu} f^{(0)} =0\,.
\end{aligned}\label{linear_gen_curved}
\end{equation}
\item higher order equations 

\begin{equation}
  \big(\tilde{\triangledown}^{\mu}-2\partial^{\mu}\tilde{\phi} \big)  \partial_{\mu}f^{(n+1)} = \sum_{p+q=n} \partial^{\mu}f^{(p)} \partial_{\mu}f^{(q)} = 0\,,
\label{higher_eom_dil_curved}\end{equation}
and
\begin{equation}
\begin{aligned}
  \tilde{\triangledown}^{-}_{\mu} \partial_{\nu} f^{(n+1)} &= \frac{1}{2} \Big( \partial^{\rho}f^{(n)} \tilde{\triangledown}_{\rho}\big(\varphi l_{\mu} \bar{l}_{\nu}\big) -\tilde{\triangledown}_{\mu}\big(\varphi l^{\rho} \bar{l}_{\nu}\partial_{\rho}f^{(n)}\big) -\tilde{\triangledown}_{\nu}\big(\varphi l_{\mu} \bar{l}^{\rho}\partial_{\rho}f^{(n)}\big)
  \\ 
  &\quad + \partial^{\rho}f^{(n)} \tilde{H}_{\rho\mu}{}^{\sigma} \varphi l_{\sigma} \bar{l}_{\nu} -\partial^{\rho}f^{(n)} \tilde{H}_{\rho\nu}{}^{\sigma} \varphi l_{\mu} \bar{l}_{\sigma} +\partial^{\rho}f^{(n)} \tilde{H}_{\mu\nu}{}^{\sigma} \varphi l_{(\rho}\bar{l}_{\sigma)}\Big)\,.
\end{aligned}\label{higher_eom_gen_curved}
\end{equation}
\end{itemize}
As in the flat background case, $\varphi$, $l^{\mu}$, $\bar{l}^{\mu}$ and $f^{(0)}$ can be completely determined from the linear order equations. The higher order equations define a set of recursion relations for $f^{(n)}$, $n>0$. Instead of the recursion relations, it is better to exploit the generalized KS equations \eqref{eom_curved_dil} and \eqref{eom_curved_gen} to determine $f$. If we plug the solution of the linear equations, $\varphi$, $l^{\mu}$ and $\bar{l}^{\mu}$, back into \eqref{eom_curved_dil} and \eqref{eom_curved_gen}, we get linear equations with respect to $f$ and $\Phi$. Hence, this shows that DFT and supergravity field equations can be solved using the linear equations only.

\section{Killing Spinor equation}
The Killing spinor equation provides a useful technique in finding solutions to the equations of motion that preserve some fraction of supersymmetry. Essentially they reduce the supergravity equations to first order in derivatives and often, when combined with a suitable ansatz for the metric and fields, will lead to linear equations. In what follows we investigate the Killing spinor equation with generalized Kerr-Schild ansatz for the 10-dimensional $N=1$ supersymmetric DFT. We will show that the Killing spinor equation is linearized under the generalized Kerr-Schild ansatz as the equations of motion. One can then seek solutions preserving fractions of supersymmetry. 

The maximal and half-maximal supersymmetric DFT are constructed in \cite{}. The field content of the fermion sector of $N=1$ supersymmetric DFT is given by the  dilatino and the gravitino
\begin{equation}
  \rho^{\alpha}\,, \qquad \psi_{\bar{p}}^{\alpha}\,,
\label{}\end{equation}
where $\alpha, \beta,\cdots$ are spinor indices for $\mathit{SO}(1,9)$ local Lorentz group. The $\mathit{SO}(1,9)$ Clifford algebra is
\begin{equation}
  \big(\gamma^{m}\big)^{*} = \gamma^{m} \,, \qquad \gamma^{m} \gamma^{n} + \gamma^{n} \gamma^{m} = 2 \eta^{mn}\,,
\label{}\end{equation}
and the chirality operator $\gamma^{(11)} = \gamma^{0}\gamma^{1}\cdots \gamma^{9}$. 
 All the spinors are taken to be Majorana-Weyl spinors with definite chiralities
\begin{equation}
  \gamma^{(11)} \rho = -\rho \,, \qquad \gamma^{(11)} \psi_{\bar{p}} = \psi_{\bar{p}}\,
\label{}\end{equation}
where $\gamma^{(11)} = \gamma^{0}\gamma^{1}\cdots \gamma^{9}$.

The SUSY variation of fermions provides the Killing spinor equations, which are 
\begin{equation}
\begin{aligned}
  \delta \rho &= -\gamma^{p} {\cal D}_{p} \varepsilon = - \gamma^{p} V_{p}{}^{M} \partial_{M} \varepsilon - \frac{1}{4} V^{M}{}_{p}\Phi_{Mmn} \gamma^{p m n} \varepsilon -\frac{1}{2} V^{Mm}\Phi_{Mmn}\gamma^{n} \varepsilon=0\,,
  \\
  \delta \psi_{\bar{p}}	&= \bar{V}^{M}{}_{\bar{p}} {\cal D}_{M} \varepsilon = \bar{V}^{M}{}_{\bar{p}} \partial_{M}\varepsilon + \frac{1}{4} \bar{V}^{M}{}_{\bar{p}} \Phi_{Mmn} \gamma^{mn} \varepsilon =0\,,
\end{aligned}\label{}
\end{equation}
where $\varepsilon^{\alpha}$ is the supersymmetry parameter which carries the same chirality as the gravitino, $\gamma^{(11)}\varepsilon= \varepsilon$.

We now want to represent the Killing spinor equations in terms of supergravity fields. Substituting the double spin connection in \eqref{DFT_Connection_GKS}, we obtain
\begin{equation}
\begin{aligned}
  &\gamma^{p}\Big[ - \frac{1}{12}\tilde{H}_{pmn}\gamma^{mn}\varepsilon  - \partial_{p} \tilde{\phi} \varepsilon + \big(\tilde{e}_{p}{}^{\mu} + \frac{1}{2} \kappa \varphi l_{p} \bar{l}^{\mu}\big) \tilde{D}^{+}_{\mu}\varepsilon + \frac{1}{4}\kappa\tilde{D}^{+}_{\bar{q}}\big(\varphi l_{p}\bar{l}^{\bar{q}}\big) \varepsilon
  \\
  & \quad\quad - \kappa \big(\tilde{e}_{p}{}^{\mu} + \frac{1}{2} \kappa \varphi l_{p} \bar{l}^{\mu}\big) \partial_{\mu} f \varepsilon \Big] = 0\,,
\end{aligned}\label{SUSY_dilatino}
\end{equation}
and 
\begin{equation}
  \big(\tilde{\bar{e}}_{\bar{p}}{}^{\mu} + \frac{1}{2} \kappa \varphi l^{\mu} \bar{l}_{\bar{p}}\big)  \tilde{D}^{+}_{\mu} \varepsilon + \frac{\kappa}{4}\tilde{D}_{m}\big(\varphi l_{n} \bar{l}_{\bar{p}}\big) \gamma^{mn} \varepsilon - \frac{\kappa}{8}\tilde{H}_{m\bar{p}\bar{q}} \big(\varphi l_{n} \bar{l}^{\bar{q}}\big) \gamma^{mn}\varepsilon = 0\,.
\label{SUSY_gravitino}\end{equation}
For simplicity, let us choose $\varepsilon$ as a Killing spinor for the background geometry satisfying
\begin{equation}
\begin{aligned}
  \partial_{p}\phi \gamma^{p} \varepsilon_{0} +\frac{1}{12} \tilde{H}_{mnp} \gamma^{mnp}\varepsilon_{0} = 0\,,
  \\
  \tilde{D}^{+}_{\bar{p}}\varepsilon_{0} = 0\,,
\end{aligned}\label{}
\end{equation}
where $\varepsilon_{0}$ is the background Killing spinor. Then the Killing spinor equations are greatly simplified as
\begin{equation}
  \frac{1}{4}\kappa\tilde{D}_{\bar{q}}\big(\varphi l_{p}\bar{l}^{\bar{q}}\big) \gamma^{p}\varepsilon_{0} + \frac{1}{8}\kappa \tilde{H}_{\bar{q}pr}\big(\varphi l^{r}\bar{l}^{\bar{q}} \big)\gamma^{p}\varepsilon_{0}  -\kappa \big(\tilde{e}_{p}{}^{\mu} + \frac{1}{2} \kappa \varphi l_{p} \bar{l}^{\mu}\big) \partial_{\mu} f \gamma^{p}\varepsilon_{0} = 0\,,
\label{}\end{equation}
and 
\begin{equation}
  \Big(\tilde{D}_{m}\big(\varphi l_{n} \bar{l}_{\bar{p}}\big) - \frac{1}{2} \tilde{H}_{m\bar{p}\bar{q}} \big(\varphi l_{n} \bar{l}^{\bar{q}}\big) \Big)\gamma^{mn}\varepsilon_{0} = 0\,. 
\label{SUSY_gravitino}\end{equation}

After converting to ten-dimensional vector indices using the pair of vielbeins, $e_{\mu}^{m}$ and $\bar{e}_{\mu}{}^{\bar{m}}$, and a field redefinition, $\Psi = e^{-2\kappa f}$ and $\varphi' = e^{-2\kappa f} \varphi$, we get completely linear equations
\begin{tcolorbox} 
\begin{equation}
  \Big(\partial_{\mu} \Psi + \frac{1}{2} \tilde{D}^{+}_{\nu}\big( \varphi' l_{\mu} \bar{l}^{\nu}\big)\Big)\gamma^{\mu} \varepsilon_{0} = 0\,,
\label{}\end{equation}
and 
\begin{equation}
  \Big(\tilde{D}_{\mu}\big(\varphi l_{\nu}\bar{l}_{\rho}\big) -\frac{1}{2} \tilde{H}_{\mu\rho\sigma} \big(\varphi  l_{\nu} \bar{l}^{\sigma}\big)\Big)\gamma^{\mu\nu} \varepsilon_{0}=0\,.
\label{gravitino_KillingSpinor}\end{equation}
\end{tcolorbox}

We see that for a flat background case the Killing spinor equations yield
\begin{equation}
\begin{aligned}
 \Big( \partial_{\mu} \Psi + \frac{1}{2} \partial_{\nu}\big(\varphi' l_{\mu}\bar{l^{\nu}}\big)\Big) \gamma^{\mu} \varepsilon_{0} = 0\,,
\end{aligned}\label{}
\end{equation}
\begin{equation}
\begin{aligned}
  \partial_{\mu} \big(\varphi l_{\nu}\bar{l}_{\rho}\big) \gamma^{\mu\nu} \varepsilon_{0} = 0\,,
\end{aligned}\label{gravitino_KillingSpinor_flat}
\end{equation}
where $\varepsilon_{0}$ is a constant spinor. These equations are remarkably simple, and much easier to solve than the full Killing spinor equations. 

\subsection{Supersymmetric classical double copy}
In section \ref{sec:3.3} we showed that the classical double copy can be extended to the entire NS-NS sector. Here we will show that the classical double copy is still valid for supersymmetric backgrounds by considering Killing spinor equations. For simplicity we shall focus on a flat background only. From the single copy relation, we have shown that a pair of Maxwell equations can be obtained from supergravity equations of motion by contracting a Killing vector with $\mathcal{R}_{\mu\nu}$. Thus it is natural to guess that we may construct BPS equation of supersymmetric Maxwell theory from the Killing spinor equation for the gravitino. 

Let us contract a Killing vector with the Killing spinor equation for the gravitino \eqref{gravitino_KillingSpinor_flat}. As before, we choose a  coordinate which makes the Killing vector constant. Then using the previous identification of the single copy relation, $A_{\mu} = \varphi l_{\mu}$, we obtain a BPS equation for supersymmetric Maxwell theory as expected
\begin{equation}
  F_{\mu\nu} \gamma^{\mu\nu} \varepsilon_0 = 0\,.
\label{BPS}\end{equation}
Since there are no more indices to contract in \eqref{BPS}, it is not possible to define a supersymmeric zeroth copy.

Recall that in section \ref{sec:3.3} we have shown a pair of gauge fields arise from the single copy relation of DFT. One may wonder why there is only a single gauge field in \eqref{BPS}. In fact, to incorporate the other gauge field to the supersymmetric single copy, we need to consider Killing spinor equation for $N=2$ supersymmetry. Since there are two different gravitinos $\psi_{\bar{m}}$ and $\psi'_{m}$, their SUSY variations provide two individual BPS equations. If we neglect the Ramond-Ramond sector, one can obtain two BPS equations as follows:
\begin{equation}
\begin{aligned}
  F_{\mu\nu} \gamma^{\mu\nu} \varepsilon_0 = 0\,, \qquad \bar{F}_{\mu\nu} \gamma^{\mu\nu} \varepsilon_0 = 0\,,
\end{aligned}\label{}
\end{equation}
where $F_{\mu\nu}$ and $\bar{F}_{\mu\nu}$ are the field strengths defined in \eqref{Field_Strength}. This shows the supersymmetric classical double copy, which states that solutions of the Killing spinor equation can be represented by solutions of the BPS equation of the supersymmetric Maxwell theory.

\section{Examples}

\subsection{Chiral null model}
We apply the previous results to the chiral null model \cite{Horowitz:1994rf} which is a class of string backgrounds that have one conserved chiral null current in the sigma-model. These are a generalization of the pp-wave and fundamental string background and have a null Killing vector and unbroken supersymmetries. Special cases are the Taub-NUT geometry and rotating black holes. The explicit geometry is given by
\begin{equation}
\begin{aligned}
  {\rm d}s^{2} &= F(x^{i}) {\rm d} u \Big({\rm d}v + K(u,x^{i}) {\rm d}u + 2 V_{i}(u,x^{i}) {\rm d}x^{i}\Big) + {\rm d}x^{i}{\rm d}x^{i}\,,
  \\
  B_{uv} &= F(x^{i})\,, \qquad B_{ui} = 2F(x^{i}) V_{i}(u,x^{i})\,,
  \\
  \phi &= \phi(u) + \frac{1}{2} \log F(x^{i}) \,,
\end{aligned}\label{CNM}
\end{equation}
where $b_{i}$ is a constant vector. We assume that the transverse space is flat, but it is possible to generalize to any conformal transverse space \cite{Horowitz:1994rf}. 

Now we show that  \eqref{CNM} fits into the generalized Kerr-Schild ansatz in a flat background. If we introduce $\tilde{v}$, $\tilde{V}_{i}$ and $X(x,u)$, such that
\begin{equation}
\begin{aligned}
  V_{i} &= \tilde{V}_{i} + \frac{1}{2} \partial_{i}X\,, \qquad v = \tilde{v} - X(x,u) \,, 
  \\
  X(x,u) &= \int^{u} \Big(K + \frac{4F}{(F-1)} \tilde{V}_{i} \tilde{V}^{i}\Big)(\vec{x},u') {\rm d}u'\,,
\end{aligned}\label{}
\end{equation}
the \eqref{CNM} is rewritten in the form
\begin{equation}
\begin{aligned}
  {\rm d}s^{2} = {\rm d}u{\rm d}\tilde{v} + {\rm d}x^{i} {\rm d}x^{i} + (F-1) {\rm d}u
\Big({\rm d}\tilde{v}- \Big( \frac{2F}{F-1}\Big)^{2}\tilde{V}_{i}\tilde{V}^{i} {\rm d}u +\frac{2F}{F-1}\tilde{V}_{i}{\rm d}x^{i}\Big)\,.
\end{aligned}\label{gKS_CNM}
\end{equation}

The associated $\varphi$ and null vectors $l$ and $\bar{l}$ can be easily read off from \eqref{genKS_gb} 
\begin{equation}
\begin{aligned}
  & \kappa\varphi = F^{-1}-1\,,
  \\
  &l_{u}= 1\,,
  \\
  &\bar{l}_{u}= 	- \Big( \frac{2F}{F-1}\Big)^{2}\tilde{V}_{i} \tilde{V}^{i}\,, \qquad \bar{l}_{\tilde{v}} = 1\,, \qquad \bar{l}_{i} = \frac{2F}{F-1}\tilde{V}_{i}\,,
\end{aligned}\label{}
\end{equation}
and one can check that $l$ and $\bar{l}$ are null vectors with respect to the flat background metric
\begin{equation}
  \eta = \begin{pmatrix} 0 & \frac{1}{2} &  0
  \\ \frac{1}{2} & 0 & 0 \\ 0 & 0 & \mathbf{1}_{d-2} \end{pmatrix}\,, \qquad  \mathbf{1}_{d-2} = \mbox{diag} (\underbrace{1,1, \cdots ,1}_{d-2})\,.
\label{}\end{equation}
It is obvious that $l_\mu$, $\bar{l}_{\mu}$ and $f$ satisfy the generalized geodesic equation \eqref{gengeo}, because $l = {\rm d}u$.

Substituting the above generalized KS ansatz into the equations of motion, we get
\begin{equation}
  \partial_{i}\partial^{i} f + \kappa \partial_{i}f \partial^{i}f = 0\,,
\label{CNM_eom_dil}\end{equation}
and the nonvanishing component of \eqref{eom_gen} are
\begin{equation}
\begin{aligned}
  &~ \partial_{i} \partial^{i} F^{-1} - 2\kappa \partial_{i}f \partial^{i}F^{-1} = 0\,,
  	\\
  &~ e^{2\kappa f}\Big(2\partial^{i}\big(e^{-2\kappa f} \partial_{u}V_{i} \big)-\partial^{i}\big(e^{-2\kappa f} \partial_{i}K \big)\Big) +4\kappa F^{-1}\partial^{2}_{u}f =0\,,
  \\
  &~ \partial_{i}\partial_{j}f =0\,,
  \\
  &~ \partial_{i}\partial_{u}f =0\,,
  \\
  &~ -4\big(\partial^{j}\mathcal{F}_{ji}-2\kappa \partial^{j} f \mathcal{F}_{ji}\big) -2 e^{2\kappa f}\partial_{u}\big(e^{-2\kappa f}\big) \partial_{i}F^{-1} = 0\,.
\end{aligned}\label{CNM_eom}
\end{equation}
where $\mathcal{F}_{ij}=\partial_{i}V_{j} - \partial_{j}V_{i}$. 
From the third and fourth equations, we can restrict $\kappa f = \phi(u)$, where $\phi(u)$ is an arbitrary function of $u$.

It is shown that if we allow a linear dilaton term, $b_{i} x^{i}$, \eqref{CNM_eom_dil} should be modified to the effective field theory for a noncritical string theory \cite{Horowitz:1994rf}. However, we shall consider only critical string theory and set $b_{i}=0$.

Substituting $\kappa f = \phi(u)$ into \eqref{CNM_eom}, we get a set of equations
\begin{equation}
\begin{aligned}
  \partial_{i} \partial^{i} F^{-1}  &= 0\,,
  \\
  -\partial^{i}\partial_{i} K +2\partial^{i}\partial_{u}V_{i} +4 F^{-1}\partial_{u}^{2}\phi &=0\,,
  \\
  -4\partial^{j}\mathcal{F}_{ji} +4 \partial_{u}\phi \partial_{i}F^{-1}&=0\,.
\end{aligned}\label{CNM_eom2}
\end{equation}
This is the same exactly with the equation (2.12) in \cite{Callan:1995hn}.

Now let us apply the strategies discussed in section \ref{sec:3.2}. If we substitute the generalized KS ansatz into \eqref{SolvEOM_flat}, we have
\begin{equation}
\begin{aligned}
  \partial^{i}\partial_{i}F^{-1} &=0\,,
  \\
  -\partial^{i}\partial_{i} K +2\partial^{i}\partial_{u}V_{i} + \partial_{u}\partial_{u} H(u,x^{i}) &=0\,,
  \\
  -4\partial^{j}\mathcal{F}_{ji}  +\partial_{u} \partial_{i}H(u,x^{i})&=0\,,
    \\
   2\partial_{i}\partial_{j}F^{-1} +\partial_{i}\partial_{j}H(u,x^{i}) &=0\,,
   \\
   \partial_{i}\partial_{u}H(u,x^{i})&=0\,.
\end{aligned}\label{linear_f0}
\end{equation}
Here $H= H(u,x^{i})$ is a harmonic function, and satisfies $\partial^{i}\partial_{i} H = 0$. From the last two equations of \eqref{linear_f0}, we determine $H(u,x^{i}) = 4\phi(u) - 2F^{-1}$.  
If we substitute the $H(u,x^{i})$ into \eqref{linear_f0} 
\begin{equation}
\begin{aligned}
  \partial^{i}\partial_{i}F^{-1} &=0\,,
  \\
  -\partial^{i}\partial_{i} K +2\partial^{i}\partial_{u}V_{i} + 4\partial_{u}\partial_{u} \phi&=0\,,
  \\
  -4\partial^{j}\mathcal{F}_{ji} &=0\,.
\end{aligned}\label{}
\end{equation}
To be consistent with \eqref{CNM_eom2}, there are two options: $\phi = 0$ or $F = \mbox{const}$. If we assume that $\phi = 0$ and $V_{i} = V_{i}(x^{i})$, then the equations of motion implies $F^{-1}$ and $K$ are harmonic function and $V_{i}$ is the $U(1)$ gauge field as discussed in \cite{Callan:1995hn, Horowitz:1994rf}. Therefore, this confirms that the linear equations are enough to solve the equations of motion.

\subsection{F1-NS5 system}
To examine our formalism in the case of a curved background in section \ref{sec 4}, let us consider superposition of a number $Q_{1}$ of fundamental strings and $Q_{5}$ of NS5 brane system. We wrap a number $Q_{5}$ of NS5-branes on $T^{5}$ along $x_{5},\cdots x_{9}$. The fundamental strings wrap one of the directions of the torus along $x^{5}$ direction.  The corresponding geometry is given by
\begin{equation}
\begin{aligned}
  {\rm d}s^{2} &= F_{1}^{-1}\big(-{\rm d}t^{2}+{\rm d}x^{2}_{5}\big) + F_{5} \big({\rm d}x_{1}^{2}+ \cdots + {\rm d} x_{4}^{2}\big) + {\rm d}x_{6}^{2}+ \cdots {\rm d}x_{9}^{2}
  \\
  e^{-2\phi} &= g_{s} F_{1} F^{-1}_{5}
  \\
  B_{05} &= F^{-1}_{1} -1\,,
  \\
  H_{ijk} &= \epsilon_{ijkl} \partial^{l}F_{5}\,, \qquad i,j,k,l = 1,2,3,4
\end{aligned}\label{}
\end{equation}
where $\epsilon_{ijkl}$ is the flat space epsilon tensor and 
\begin{equation}
  F_{1}=1+\frac{16\pi^{4} \alpha'^{3}Q_{1}}{g_{s}^{2}V_{4} r^{2}}\,, \qquad F_{5} = 1 + \frac{\alpha' Q_{5}}{r^{2}}\,,
\label{}\end{equation}
here $V_{4}$ is the volume of the $T^{4}$. 
This geometry is the S-dual of the D1-D5 system, and it is a particular case of the chiral null model with the conformal transverse background. In this case,  $B_{ui}$ does not have to be the same with $g_{ui}$. 

The NS5-brane background is treated as a background
\begin{equation}
\begin{aligned}
    {\rm d}\tilde{s}^{2} &= -{\rm d}t^{2} + F_{5} \big({\rm d}x_{1}^{2}+ \cdots + {\rm d} x_{4}^{2}\big) + {\rm d}x_{5}^{2}+ \cdots {\rm d}x_{9}^{2}
  \\
  e^{-2\tilde{\phi}} &= g_{s} F^{-1}_{5}
  \\
  \tilde{H}_{ijk} &= \epsilon_{ijkl} \partial^{l}F_{5}\,, \qquad i,j,k,l = 1,2,3,4
\end{aligned}\label{background_geom}
\end{equation}
where we put tilde for all the background quantities. 

Then using \eqref{genKS_gb}, we can read off the corresponding generalized KS ansatz 
\begin{equation}
\begin{aligned}
  \varphi = F_{1} -1	\,, \qquad l = {\rm d} t + {\rm d} x_{5}\,, \qquad \bar{l} = - {\rm d} t + {\rm d}x_{5}\,.
\end{aligned}\label{gKS_ansatz_ex2}
\end{equation}
Note that $l_{\mu}$ and $\bar{l}_{\mu}$ are orthogonal to the background 3-form flux, $\tilde{H}_{\mu\nu\rho}$
\begin{equation}
  l^{\mu} \tilde{H}_{\mu\nu\rho} = \bar{l}^{\mu} \tilde{H}_{\mu\nu\rho}=0\,,
\label{ortho_H}\end{equation}
and $\phi = \frac{1}{\sqrt{-g}}$ of
\begin{equation}
  f=0\,.
\label{f=0}\end{equation}
Under the \eqref{ortho_H} and \eqref{f=0}, the generalized KS field equations \eqref{eom_curved_dil} and \eqref{eom_curved_gen} are greatly simplified
\begin{equation}
\begin{aligned}
  &\kappa\Big[\ \big(\tilde{\triangledown}_{\mu} -2\partial_{\mu}\tilde{\phi}\big)\big(\tilde{\triangledown}_{\nu} -2\partial_{\nu}\tilde{\phi} \big)\big(\varphi l^{(\mu}\bar{l}^{\nu)}\big) \Big] =0\,,
  \\
  &\frac{\kappa}{4} \Big[\ \big({\tilde{\triangledown}}_{\rho}-2 \partial_{\rho}\tilde{\phi}\big){\tilde{\triangledown}}^{\rho}\big(\varphi l_{\mu} \bar{l}_{\nu}\big) -\big({\tilde{\triangledown}}^{\rho}-2\partial^{\rho}\tilde{\phi}\big) {\tilde{\triangledown}}_{\mu}\big(\varphi l_{\rho}\bar{l}_{\nu}\big) 
- \big({\tilde{\triangledown}}^{\rho}-2\partial^{\rho}\tilde{\phi}\big) {\tilde{\triangledown}}_{\nu}\big(\varphi l_{\mu}\bar{l}_{\rho}\big)\Big]= 0\,,
\end{aligned}\label{}
\end{equation}
where $\tilde{\triangledown}_{\mu}$ is the covariant derivative for the background metric \eqref{background_geom}. Thus these equations are completely linear and it is straightforward to show that the generalized KS ansatz \eqref{gKS_ansatz_ex2} satisfies the generalized KS field equations for the curved backgrounds.

\subsection{Charged black string}
It is well-known that momenta and axion charges are interchangeable  via T-duality \cite{Horne:1991cn}. For simplicity consider a 5-dimensional (uncharged) black string and denote the longitudinal direction along the string as $y$. Then the charged black string solution can be obtained from the black string solution by boosting along the $y$ direction, which generates an off-diagonal component of the metric, and then taking T-duality along $y$ direction. The explicit geometry is given by
\begin{equation}
\begin{aligned}
 {\rm d}s^{2} &= \Big(1+\frac{2m S^{2}}{r}\Big)^{-1}\Big[-\Big(1-\frac{2m}{r}\Big){\rm d}t^{2} + {\rm d}y^{2} \Big]+ \Big(1-\frac{2m}{r}\Big)^{-1} {\rm d}r^{2} +r^{2}{\rm d}\Omega \,,
 \\
 B_{yt} &= \frac{C}{S}\Big(1+\frac{2mS^{2}}{r}\Big)^{-1}\,, \qquad\qquad e^{-2\phi} = 1+\frac{2mS^{2}}{r}\,,
\end{aligned}\label{CBS}
\end{equation}
where $C=\cosh \alpha$ and $S= \sinh \alpha$, and $\alpha$ is a boost parameter. 

Let us start from the black string solution. It is a direct product of Schwarzschild black hole and a circle $S^{1}$
\begin{equation}
\begin{aligned}
  {\rm d}s^{2} &= - \Big(1-\frac{2M}{r}\Big){\rm d}\tilde{t}^{2} + \Big(1-\frac{2M}{r}\Big)^{-1}{\rm d}r^{2} + r^{2}{\rm d}\Omega^{2} + {\rm d}y^{2}\,.
\end{aligned}\label{}
\end{equation}
where $r$ is transverse radius.
Using the conventional Kerr-Schild ansatz for Schwarzschild BH in Eddington-Finkelstein coordinate, we have
\begin{equation}
  {\rm d}s^{2} = -{\rm d}\hat{t}^{2} + {\rm d}r^{2} + r^{2}{\rm d}\Omega^{2} + {\rm d}y^{2} -\kappa\varphi \big({\rm d}\hat{t} + {\rm d}r\big)^{2}\,, \qquad \kappa \varphi = -\frac{2M}{r}\,,
\label{}\end{equation}
where $\hat{t} = \tilde{t} + \big(r_{*}-r\big)$ and $r_{*}$ is the tortoise coordinate defined
\begin{equation}
  r_{*} = r +2M \log\Big|\frac{r}{2M} -1\Big|\,,\qquad {\rm d}r_{*} = {\rm d}r\Big(1-\frac{2M}{r}\Big)^{-1}\,.
\label{}\end{equation}

Applying a boost along the $y$-direction, $\hat{t}\to \hat{t} \cosh \alpha + y \sinh \alpha $, $y \to \hat{t} \sinh \alpha + y \cosh \alpha$, we get
\begin{equation}
  {\rm d}s^{2} = -{\rm d}\hat{t}^{2} + {\rm d}r^{2} + r^{2}{\rm d}\Omega^{2} + {\rm d}y^{2} -\kappa\varphi \big({C \rm d}\hat{t} + S{\rm d}y + {\rm d} r\big)^{2}\,,
\label{}\end{equation}
and the null vectors are identical, $l = \bar{l} = C {\rm d} \hat{t} + S {\rm d}y + {\rm d}r$. 

Let us take a T-duality along the $y$-direction. According to the Buscher's rule for the null vectors \eqref{Buscher_null}, the null vectors split into
\begin{equation}
  l \to l' = C {\rm d} \hat{t} + S {\rm d}y + {\rm d}r\,, \qquad \bar{l} \to \bar{l}' = C {\rm d} \hat{t} - S {\rm d}y + {\rm d}r \,,
\label{}\end{equation}
and $l\cdot \bar{l} = -2 S^{2}$. The corresponding generalized metric and Kalb-Ramond field are 
\begin{equation}
\begin{aligned}
  {\rm d}s'^{2} &= \Big(1+\frac{2MS^{2}}{r}\Big)^{-1}\Big[-\Big(1-\frac{2M}{r}\Big){\rm d}\hat{t}^{2}+\frac{4MC}{r}{\rm d}\hat{t}{\rm d}r+\Big(1+\frac{2MC^{2}}{r}\Big){\rm d}r^{2}+{\rm d}y^{2}\Big] + r^{2} {\rm d}\Omega^{2}
  \\
  B'&= \Big(-\frac{C}{S} + \frac{2MCS}{r}\Big(1+\frac{2MS^{2}}{r}\Big)^{-1}\Big) {\rm d} \hat{t}\wedge {\rm d}y -\frac{2MS}{r}\Big(1+\frac{2MS^{2}}{r}\Big)^{-1}{\rm d}r\wedge {\rm d}y\,.
\end{aligned}\label{}
\end{equation}
Here we have added $-\frac{C}{S}$ to the $B'{}_{\hat{t}y}$, which does not contribute to the 3-form field strength. 

Finally, we make a further coordinate transform $\hat{t} = t+ C\big(r_{\ast} -r\big)$ and ignore the $B'(r)_{ry}$ component which does not contribute to the field strength, we get the previous charged black string geometry \eqref{CBS}. Thus \eqref{CBS} is equivalent to the following generalized KS ansatz:
\begin{equation}
\begin{aligned}
  {\rm d}s^{2} &= -{\rm d}\hat{t}^{2} + {\rm d}r^{2} + r^{2}{\rm d}\Omega^{2} + {\rm d}y^{2} + \frac{2M}{r+2MS^{2}}\big(C {\rm d} \hat{t} + S {\rm d}y + {\rm d}r\big)\big(C {\rm d} \hat{t} - S {\rm d}y + {\rm d}r\big)\,,
  \\
  B'&= \frac{2M}{r+2MS^{2}} \big(C {\rm d} \hat{t} + S {\rm d}y + {\rm d}r\big)\wedge \big(C {\rm d} \hat{t} - S {\rm d}y + {\rm d}r\big)\,.    
\end{aligned}\label{}
\end{equation}
In this example, the DFT dilaton vanishes, thus the generalized KS field equations become linear.


\section{Summary}
In this paper we have generalized the conventional KS formalism to DFT and the corresponding supergravities. We have introduced the generalized Kerr-Schild ansatz for the generalized metric for DFT and represented it in terms of supergravity fields.  The key ingredient is the pair of null vectors. We showed by means of the null vectors that the generalized KS ansatz satisfies the $\mathit{O}(d,d)$ constraint automatically. The generalized KS ansatz for the generalized metric is linear in the expansion parameter $\kappa$, however, the corresponding metric and Kalb-Ramond fields are nonlinear in $\kappa$. We have analyzed how Buscher's transform acts on the generalized KS ansatz and showed that T-duality maps a generalized KS ansatz to another generalized KS ansatz. 

The field equations for the generalized KS ansatz have been constructed from the DFT field equations. The consistency condition requires that the null vectors satisfy the generalized geodesic condition, which reduces to the conventional geodesic condition when the two null vectors are equal. Even though the field equations are quadratic in $\kappa$, it is possible to solve the field equations by using the linear equations only. We showed that the field equations can be always reduced to the linear equations by using the series expansion of the DFT dilaton. It is remarkable that the supergravity equations of motion can be solved by considering the linear equations only. 

Based on the generalized KS formalism, we have shown how to extend the classical double copy to the entire NS-NS sector. A pair of Maxwell equations are derived from the generalized KS equation by contracting a Killing vector with the generalized KS field equation, and each null vector is identified with different gauge fields. We have further shown that the BPS equation of the supersymmetric Maxwell theory can be realized from the Killing spinor equation by contracting a Killing vector with the Killing spinor equation for the gravitino. 

We have considered pure DFT, which consists of the massless NS-NS sector only, but in general we can incorporate additional matter fields, such as Yang-Mills gauge fields, the Ramond-Ramond field, fermion fields and various brane sources in string theory \cite{Angus:2018mep}. The present formalism can be extended to heterotic DFT and various M-theory extensions \cite{MDFT1,MDFT2,MDFT3,MDFT4,MDFT5,MDFT6}. To this end, we need to introduce a generalized KS ansatz for the generalized metric for each theory. Especially, the generalization to heterotic DFT \cite{Hohm:2011ex} is straightforward, because the structure is almost parallel with conventional DFT. In this case we do not have to consider additional matter fields, including higher derivative corrections \cite{Bedoya:2014pma,Coimbra:2014qaa,Lee:2015kba}, since all the field contents are encoded in the generalized metric. Furthermore, it is also interesting to consider the classical double copy for heterotic supergravity. 

The other straightforward generalization is a dimensional reduction of DFT and supergravities to lower dimensions. For the DFT/EFT side, various compactifications, such as Kaluza-Klein reduction or the generalized Scherk-Schwarz compactification \cite{GSS1,GSS2,GSS3,GSS4,GSS5,GSS6,Cho:2015lha}, have been studied extensively, and it is possible to apply these results to develop the generalized KS formalism for lower dimensional DFT and supergravities. It would also be interesting to consider the classical double copy for the lower dimensional theories.

\section*{Acknowledgement}
We thank Jeong-Hyuck Park, Alejandro Rosabal, Vladislav Vaganov and Sang-Heon Yi for useful comments and suggestions. We would also like to thank the organizers and participants of the workshop ``100+3 General Relativity Meeting'' at Jeju National University where the idea for this paper were shown for the first time.


\newpage
\appendix
\section{Conventional Kerr-Schild formalism}\label{ReviewKS}
The conventional Kerr-Schild metric ansatz is given by
\begin{equation}
  g_{\mu\nu} = \tilde{g}_{\mu\nu} + \kappa\phi \ell_{\mu}\ell_{\nu}
\label{KerrShild}\end{equation}
where $\tilde{g}$ is a background metric satisfying the vacuum Einstein equation, and $\ell$ is a null vector with respect to both $g$ and $\tilde{g}$
\begin{equation}
  g^{\mu\nu}\ell_{\mu} \ell_{\nu} = \tilde{g}^{\mu\nu}\ell_{\mu} \ell_{\nu} = 0\,.
\label{invKerrSchild}\end{equation}
The virtue of this ansatz is that the inverse metric $g^{\mu\nu}$ is the same as the linearized perturbation,
\begin{equation}
  g^{\mu\nu} = \tilde{g}^{\mu\nu} - \kappa \phi \ell^{\mu} \ell^{\nu}\,,
\label{}\end{equation}
and the determinant is unchanged 
\begin{equation}
  \det(g_{\mu\nu}) = \det(\tilde{g}_{\mu\nu})\,.
\label{}\end{equation}
The indices of $\ell$ are raised and lowered by both $g$ and $\tilde{g}$ due to the null property
\begin{equation}
  \ell^{\mu} = g^{\mu\nu} \ell_{\nu} = \tilde{g}^{\mu\nu} \ell_{\nu}\,.
\label{}\end{equation}

One can show that the Christoffel connection $\gamma^{\mu}_{\nu\rho}$ for the Kerr-Shild ansatz \eqref{KerrShild} satisfies the following conditions:
\begin{equation}
  \gamma^{\mu}_{\nu\rho} \ell^{\nu} \ell^{\rho} = \tilde{\gamma}^{\mu}_{\nu\rho} \ell^{\nu} \ell^{\rho}\,, \qquad \gamma^{\mu}_{\nu\rho} \ell_{\mu}\ell^{\rho} = \tilde{\gamma}^{\mu}_{\nu\rho} \ell_{\mu}\ell^{\rho} \,.
\label{}\end{equation}
Consider a vacuum spacetime, which satisfies the vacuum Einstein equation, $R_{\mu\nu}= 0$. As a consistency relation, the contraction of $\ell^{\mu}$ with $R_{\mu\nu}$ should vanish,
\begin{equation}
   R_{\mu\nu} \ell^{\mu}\ell^{\nu} = -\kappa \phi g^{\nu\sigma} \big(\ell^{\mu}\tilde{\triangledown}_{\mu}\ell_{\nu}  \big) \big(\ell^{\rho}\tilde{\triangledown}_{\rho}\ell_{\sigma}  \big)=0\,,
\label{vacuumEinstein}\end{equation}
where $\tilde{\triangledown}_{\mu}$ is the covariant derivative with respect to the background metric $\tilde{g}$.
Assume that both the background $\tilde{g}$ and the full metric $g$ satisfy the vacuum Einstein equation, $R_{\mu\nu}=\tilde{R}_{\mu\nu}=0$, where $R_{\mu\nu}$ and $\tilde{R}_{\mu\nu}$ are Ricci tensors for the full metric and background metric respectively. Then (\ref{vacuumEinstein}) vanishes and $\ell$ must satisfy the geodesic equation
\begin{equation}
  \ell^{\mu} \tilde{\triangledown}_{\mu}\ell_{\nu}= \ell^{\mu}\partial_{\mu}\ell_{\nu} = 0\,.
\label{geodesic}\end{equation}

If we assume the $\ell^{\mu}$ is null and geodesic, the vacuum Einstein equation reads
\begin{equation}
  R_{\mu\nu} = \kappa R^{(1)}{}_{\mu\nu} + \kappa^{2} \phi \ell_{\mu} \ell^{\rho}R^{(1)}{}_{\rho\nu} = 0\,,
\label{Ricci0}\end{equation}
where $R^{(1)}$ denotes the linear terms in $\kappa$,
\begin{equation}
  R^{(1)}{}_{\mu\nu} = \kappa\tilde{\triangledown}_{\rho}\Big(\tilde{\triangledown}_{(\mu}\big(\phi\ell_{\nu)} \ell^{\rho}\big) - \frac{1}{2} \tilde{\triangledown}^{\rho}\big(\phi\ell_{\mu}\ell_{\nu}\big)\Big)\,,
\label{Ricci1}\end{equation}
Hence the Einstein equation is equivalent to $R^{(1)}{}_{\mu\nu}=0$.

\section{Semi-Covariant formalism for DFT}\label{App.1}
In this appendix, we briefly review the basics of DFT and the semi-covariant formalism for DFT \cite{DFTGeom3}. 
\subsection{Double vielbein}
The field content of DFT consists of the generalized metric $\mathcal{H}_{MN}$ and the DFT dilaton $d$. The generalized metric $\mathcal{H}$ is an $\mathit{O}(d,d)$ element and satisfies the constraint $\mathcal{J}_{MN} = \mathcal{H}_{MP} \mathcal{J}^{PQ} \mathcal{H}_{QN}$, where $\mathcal{J}$ is $\mathit{O}(d,d)$ metric. To be explicit, one can solve the $\mathit{O}(d,d)$ constraint in terms of the metric $g_{\mu\nu}$ and Kalb-Ramond field $B_{\mu\nu}$ as follows:
\begin{equation}
  {\cal H}_{MN} = \begin{pmatrix} g^{\mu\nu} & - g^{\mu\rho} B_{\rho\nu} \\ B_{\mu\rho} g^{\rho\nu} & g_{\mu\nu} - B_{\mu\rho} g^{\rho\sigma} B_{\sigma\nu} \end{pmatrix}\,.
\label{paraH}\end{equation}
The DFT dilaton is a scalar density which is represented by
\begin{equation}
  e^{-2d} = \sqrt{-g} e^{-2\phi} 
\label{}\end{equation}
where $\phi$ is the supergravity dilaton. 

The local structure group of DFT is given by the maximal compact subgroup of $\mathit{O}(d,d)$ possessing the Lorentz group $\mathit{O}(1,d-1)$
\begin{equation}
  \mathit{O}(1,d-1)_{L} \times \mathit{O}(1,d-1)_{R} \subset \mathit{O}(d,d)\,.
\label{}\end{equation}
We introduce a pair of local orthonormal frame $\{V_{M}{}^{m}, \bar{V}_{M}{}^{\bar{m}}\}$ corresponding to the $ \mathit{O}(1,d-1)_{L}$ and $\mathit{O}(1,d-1)_{R}$ respectively. They satisfy the following defining properties
\begin{equation}
\begin{aligned}
  V &= PV\,,& \qquad V \eta^{-1} V^{t} &= P\,,& \qquad V^{t} {\cal J} V &= \eta \,, &
  \\
  \bar{V} &= \bar{P}\bar{V}\,,& \qquad \bar{V} \bar{\eta}^{-1} \bar{V}^{t} &= - \bar{P}\,,& \qquad \bar{V}^{t} {\cal J} \bar{V} &= -\bar{\eta} \,,&
\end{aligned}\label{defV}
\end{equation}
and are orthogonal to each other
\begin{equation}
  (V^{t})_{m}{}^{M} \bar{V}_{Mn} = 0\,,
\label{}\end{equation}
where $\eta_{mn}$ and $\bar{\eta}_{\bar{m}\bar{n}}$ are $O(1,D-1)_{L}$ and $O(1,D-1)_{R}$ metric tensors respectively.

An important step to identify DFT with supergravity is to fix a parametrization of the double-vielbeins in terms of supergravity fields. The above constraints can be solved explicitly by assuming the upper half blocks are non-degenerate, and then the  double-vielbeins are parametrized as
\begin{equation}
\begin{aligned}
  V_{M}{}^{m} = \frac{1}{\sqrt{2}}\begin{pmatrix} (e^{-1})^{m\mu} \\ (Be^{-1}+e)_{\mu}{}^{m}\end{pmatrix}\,, \qquad \bar{V}_{M}{}^{\bar{m}} = \frac{1}{\sqrt{2}}\begin{pmatrix} (\bar{e}^{-1})^{\bar{m}\mu} \\ (B \bar{e}^{-1}-\bar{e})_{\mu}{}^{\bar{m}}\end{pmatrix}\,,
\end{aligned}\label{paraV}
\end{equation}
where $e$ and $\bar{e}$ are two copies of the $d$-dimensional vielbein corresponding to the same spacetime metric 
\begin{equation}
  e_{\mu}{}^{m} e_{\nu m} = \bar{e}_{\nu}{}^{\bar{m}} \bar{e}_{\mu \bar{m}} = g_{\mu\nu}\,.
\label{}\end{equation}
One can check these relations by comparing \eqref{paraH} and \eqref{paraV}.

The $\mathit{O}(d,d)$ metric ${\cal J}_{MN}$ and the generalized metric ${\cal H}_{MN}$ can be written in terms of the double vielbein 
\begin{equation}
\begin{aligned}
  {\cal J}_{MN} &= V_{Mm} \eta^{mn} V^{t}{}_{mN} - \bar{V}_{M \bar{m}} \bar{\eta}^{\bar{m}\bar{n}}\bar{V}^{t}{}_{\bar{n}N}\,,
  \\
  {\cal H}_{MN} &= V_{Mm} \eta^{mn} V^{t}{}_{mN} + \bar{V}_{M \bar{m}} \bar{\eta}^{\bar{m}\bar{n}}\bar{V}^{t}{}_{\bar{n}N}\,.
\end{aligned}\label{cond_DV1}
\end{equation}

\subsection{Connection and Curvature}
The gauge symmetry for DFT is given by the generalized Lie derivative $\hat{\cal L}_{X}$ which is defined by
\begin{equation}
\begin{aligned}
  (\hat{\cal L}_{X} V )^{M}{}_{N} &= X^{P} \partial_{P} V^{M}{}_{N} + (\partial^{M} X_{P} - \partial_{P} X^{M}) V^{P}{}_{N} + (\partial_{N} X^{P} - \partial^{P} X_{N}) V^{M}{}_{P}\,,
\\
\hat{\cal L}_{X} d &= X^{M} \partial_{M} d - \frac{1}{2} \partial_{M} X^{M}\,,
\end{aligned}\label{genLie}
\end{equation}
The parameter $X^{M}$ consists of the ordinary diffeomorphism parameter $\xi^{\mu}$ and the one-form gauge parameter $\Lambda_{\mu}$ for $B_{\mu\nu}$ in an $\mathit{O}(d,d)$ covariant way
\begin{equation}
  X^{M} = \{\xi^{\mu}\,, \Lambda_{\mu}\}\,.
\label{}\end{equation}
Closure and Jacobi identity of the generalized Lie derivative requires the section condition
\begin{equation}
  \partial_{M} \partial^{M} \mathcal{F}_{1} = 0 \,, \qquad \partial_{M}\mathcal{F}_{1} \partial^{M} \mathcal{F}_{2} = 0\,,
\label{}\end{equation}
where $\mathcal{F}_{1}$ and $\mathcal{F}_{2}$ are arbitrary functions on doubled space. 

As for the covariant differential operator of the generalized Lie derivative (\ref{genLie}), we present a covariant derivative which can be applied to any arbitrary $\mathit{O}(d,d)$, $\mathit{O}(1,{D-1})_{L}$ and $\mathit{O}(1,{D-1})_{R}$ representation as follows 
\begin{equation}
  {\cal D}_{M} := \partial_{M}  + \Gamma_{M} + \Phi_{M} + \bar{\Phi}_{M}\, . 
\label{MasterDerivative}\end{equation}
where $\Phi_{M mn}$ and $\bar{\Phi}_{M\bar{m}\bar{n}}$ are double spin connections and $\Gamma_{MNP}$ is the semi-covariant DFT connection \cite{DFTGeom2},
\begin{equation}
\begin{aligned}
  \Gamma_{PMN} = & 2(P\partial_{P} P \bar{P} )_{[MN]} 
	+ 2 (\bar{P}_{[M}{}^{Q} \bar{P}_{N]}{}^{R} 
	- P_{[M}{}^{Q} P_{N]}{}^{R} ) \partial_{Q} P_{R P} 
\\& 
	- \frac{4}{D-1} \big(\bar{P}_{P[M} \bar{P}_{N]}{}^{Q} + P_{P[M} P_{N]}{}^{Q}) 
		\big(\partial_{Q}d + (P\partial^{R} P \bar{P})_{[RQ]}\big)\,,
\label{conn}\end{aligned}\end{equation}

The spin connections are defined by using the semi-covariant derivative
\begin{equation}
\begin{aligned}
  \Phi_{M mn} &= V^{N}{}_{m} \nabla_{M} V_{N n}= V^{N}{}_{m} \partial_{M} V_{N n} + \Gamma_{M NP} V^{N}{}_{m} V^{P}{}_{n} \,, 
  \\
  \bar{\Phi}_{M \bar{m}\bar{n}} &= \bar{V}^{N}{}_{\bar{m}} \nabla_{M} \bar{V}_{N \bar{n}}= \bar{V}^{N}{}_{\bar{m}} \partial_{M} \bar{V}_{N \bar{n}} + \Gamma_{M NP} \bar{V}^{N}{}_{\bar{m}} V^{P}{}_{\bar{m}}\,.
\label{defspinconn}\end{aligned}
\end{equation}
Although these are not gauge covariant, we can form covariant quantities by contracting double-vielbeins 
\begin{equation}
\begin{aligned}
  &\bar{V}^{M}{}_{\bar{p}}\Phi_{Mmn} = \frac{1}{\sqrt{2}}\Phi_{\bar{p}mn}\,,\qquad \Phi_{[pmn]}\,,\qquad \Phi^{p}{}_{pm}\,,
\\
& V^{M}{}_{p}\Phi_{M\bar{m}\bar{n}} = \frac{1}{\sqrt{2}}\Phi_{p\bar{m}\bar{n}} \,,\qquad \bar{\Phi}_{[\bar{p}\bar{m}\bar{n}]}\,,\qquad \bar{\Phi}^{\bar{p}}{}_{\bar{p}\bar{m}}\,.
\end{aligned}\label{covspinconnections}
\end{equation}
It is possible to represent the covariant components of the DFT spin connection in terms of double-vielbeins
\begin{equation}
\begin{aligned}
  \bar{V}^{M}{}_{\bar{p}}\Phi_{Mmn} &= \bar{V}^{M}{}_{\bar{p}} V^{N}{}_{m} \partial_{M} V_{Nn} +  V^{M}{}_{m} V^{N}{}_{n} \partial_{M} \bar{V}_{N}{}_{\bar{p}} - V^{M}{}_{n}V^{N}{}_{m} \partial_{M} \bar{V}_{N}{}_{\bar{p}}\,,
  \\
  V^{M}{}_{p}\bar{\Phi}_{M\bar{m}\bar{n}} &= V^{M}{}_{p} \bar{V}^{N}{}_{\bar{m}} \partial_{M} \bar{V}_{N\bar{n}} + \bar{V}^{M}{}_{\bar{m}} \bar{V}^{N}{}_{\bar{n}}\partial_{M}V_{Np} - \bar{V}^{M}{}_{\bar{n}} \bar{V}^{N}{}_{\bar{m}}\partial_{M}V_{Np}\,,
  \\
  V^{M}{}_{[p}\Phi_{M mn]} &= V^{M}{}_{[m} V^{N}{}_{n} \partial_{|M} V_{N|p]}\,,
  \\
  \bar{V}^{M}{}_{[\bar{p}}\bar{\Phi}_{M\bar{m}\bar{n}]} &= \bar{V}^{M}{}_{[\bar{m}} \bar{V}^{N}{}_{\bar{n}} \partial_{|M} \bar{V}_{N|\bar{p}]}\,,
  \\
  V^{Mm}\Phi_{Mmn} &= \partial^{M} V_{M n} -2 V^{M}_{n}\partial_{M}d\,,
  \\
  \bar{V}^{M\bar{m}}\bar{\Phi}_{M\bar{m}\bar{n}} &= -\partial^{M}\bar{V}_{M\bar{n}} +2\bar{V}^{M}{}_{\bar{n}}\partial_{M}d\,.
\end{aligned}\label{}
\end{equation}
They can be represented in terms of supergravity fields
\begin{equation}
\begin{aligned}
  \Phi_{\bar{p}mn} &= \bar{e}^{\mu}_{\bar{p}} \omega_{\mu mn} + \frac{1}{2} H_{\bar{p}mn} := \bar{e}^{\mu}{}_{\bar{p}} \omega^{+}{}_{\mu mn}\,,
  \\
  \Phi_{[pmn]} &= \omega_{\mu [mn}e^{\mu}{}_{p]} + \frac{1}{6} H_{pmn} :=  {\omega}^{+}{}_{\mu [mn}e^{\mu}{}_{p]} - \frac{1}{3}H_{pmn} \,,
  \\
  \Phi^{p}{}_{pn} &= e^{\mu p} \omega_{\mu p n} -2 \partial_{n}\phi 
  \\
  \bar{\Phi}_{p\bar{m}\bar{n}} &= - e^{\mu}_{p} \bar{\omega}_{\mu \bar{m}\bar{n}} + \frac{1}{2} H_{p\bar{m}\bar{n}} := - e^{\mu}{}_{p} \bar{\omega}^{-}{}_{\mu \bar{m}\bar{n}}\,,
  \\
  \bar{\Phi}_{[\bar{p}\bar{m}\bar{n}]} &= -\bar{\omega}_{\mu [\bar{m}\bar{n}}e^{\mu}{}_{\bar{p}]} + \frac{1}{6} H_{\bar{p}\bar{m}\bar{n}} :=  - \bar{{\omega}}^{-}{}_{\mu [mn}\bar{e}^{\mu}{}_{\bar{p}]}-\frac{1}{3}H_{\bar{p}\bar{m}\bar{n}}\,,
  \\
  \bar{\Phi}^{\bar{p}}{}_{\bar{p}\bar{n}} &= -\bar{e}^{\mu \bar{p}} \bar{\omega}_{\mu \bar{p} \bar{n}} +2 \partial_{\bar{n}}\phi\,.
\end{aligned}\label{rep_DFTconnection}
\end{equation}
\begin{equation}
  \partial_{\mu} d_{0} = \partial_{\mu} \tilde{\phi} - \frac{1}{2} \tilde{\gamma}^{\nu}_{\nu\mu}
\label{}\end{equation}
where $\tilde{\gamma}$ is the background Christoffel connection in GR. 

Let us turn to the semi-covariant curvature tensor $S_{MNPQ}$ which is defined
\begin{equation}
  S_{MNPQ} = \frac{1}{2} \big(R_{MNPQ} + R_{PQMN} - \Gamma^{R}{}_{MN} \Gamma_{RPQ} \big)\,,
\label{curvature}\end{equation}
where $R_{MNPQ}$ is defined from the standard commutator of the covariant derivatives
\begin{equation}
  R_{MNPQ} = \partial_{M}\Gamma_{NPQ} - \partial_{N}\Gamma_{MPQ} + \Gamma_{MP}{}^{R} \Gamma_{NRQ} -  \Gamma_{NP}{}^{R} \Gamma_{MRQ} \,.
\label{}\end{equation}
We can represent $S_{MNPQ}$ in terms of DFT spin connections
\begin{equation}
  R_{MNPQ} = F_{PQ mn} V_{M}{}^{m} V_{N}{}^{n} + \bar{F}_{PQ \bar{m} \bar{n}} \bar{V}_{M}{}^{\bar{m}} \bar{V}_{N}{}^{\bar{n}}
\label{}\end{equation}
where
\begin{equation}
\begin{aligned}
 F_{ MNmn} &= \partial _ {M} \Phi _ {Nmn} - \partial_{N} \Phi_{M mn} + \Phi_{Mm}{}^{p} \Phi_{Npn}  - \Phi _ {Nm}{}^{p} \Phi_{Mpn} 
 \\ 
 \bar{F}_ {MN \bar{m}\bar{n}} &= \partial_{M} \bar{ \Phi }_{ N \bar{m}\bar{n} } -\partial_{N} \bar{\Phi}_{M \bar{m}\bar{n}} -\bar{\Phi}_{M\bar{m}}{}^{\bar{p}} \bar{\Phi}_{N\bar{p}\bar{n}} + \bar{\Phi}_{N\bar{m}}{}^{\bar{p}} \bar{\Phi}_{M\bar{p}\bar{n}}
\end{aligned}
\end{equation}
Even though $S_{MNPQ}$ is not a tensor with respect to the generalized diffeomorphism, we can obtain tensors by contracting $S_{MNPQ}$ with the projection operators. The generalized Ricci scalar and tensor are defined
\begin{equation}
  S_{MN} = P_{M}{}^{P} \bar{P}_{N}{}^{Q} P^{RS}S_{RPSQ} ,\qquad   S := 2 P^{MN} P^{PQ} S_{MPNQ}\,,
\label{}\end{equation}
and one can show that these are covariant under the $\mathit{O}(d,d)$ and generalized diffeomorphism. We can represent them in terms of the DFT spin connections
\begin{equation}
\begin{aligned}
  S &=  4 \partial^{m} \Phi^{n}{}_{mn} - 2\Phi^{m}{}_{m}{}^{p} \Phi^{n}{}_{np} -3 \Phi^{[mnp]} \Phi_{mnp} +                                                                        \Phi^{\bar{p} mn} \Phi_{\bar{p} mn} \,,
  \\
  S_{m\bar{n}} & =-\frac {1}{2} \left( \partial_{m} \bar{\Phi}^{\bar{p}}{}_ { \bar{n} \bar{p} } -\partial^{\bar{p}} \bar{\Phi}_{ m \bar{n}\bar{p} } - \bar{\Phi}_{ m\bar{n}}{}^{\bar{p}} \bar{\Phi}^{\bar{q}}{}_{\bar{p}\bar{q}} -\Phi^{\bar{p}}{}_{ q m } \bar{\Phi}^{ q }{}_{\bar{p}\bar{n}} \right)\,.
\end{aligned}\label{generalizedcurvaturescalar}
\end{equation}
where $\partial_{m} = V_{m}{}^{M}\partial_{M}$ and  $\partial_{\bar{m}} = \bar{V}_{\bar{m}}{}^{M}\partial_{M}$\,.


\end{document}